# The unified maximum a posteriori (MAP) framework for neuronal system identification

Michael C.-K. Wu*[1,2], Fatma Deniz*[3,4,5], Ryan J. Prenger[6] and Jack L. Gallant[4,7]

[1]PROS San Francisco, CA 94607, USA

[2]Biophysics Program, University of California, Berkeley, CA 94720, USA

[3]Berkeley Institute for Data Science, University of California, Berkeley, CA 94720, USA

[4]Helen Wills Neuroscience Institute, University of California, Berkeley, CA 94720, USA

[5]International Computer Science Institute, Berkeley, CA 94704, USA

[6]Department of Physics, University of California, Berkeley, CA 94720, USA

[7]Department of Psychology, University of California, Berkeley, CA 94720, USA

Correspondence should be addressed to Jack L. Gallant, University of California at Berkeley, Department of Psychology, 2121 Berkeley Way, Berkeley, CA 94720. E-mail: gallant@berkeley.edu.

# 1 Introduction

The functional relationship between an input and a sensory neuron's response can be described by the neuron's stimulus-response mapping function, $f$. A general approach for characterizing the stimulus-response mapping function is called *system identification* (Marmarelis, 2004). Many different names have been used for the stimulus-response mapping function: kernel or transfer function (Cook and Maunsell, 2004; Marmarelis and Marmarelis, 1978; Smyth et al., 2003; Willmore and Smyth, 2003), transducer (Ringach and Shapley, 2004; Victor and Shapley, 1980), spatiotemporal (or spectrotemporal) receptive field [STRF, (Aertsen and Johannesma, 1981a; David et al., 2004; DeAngelis et al., 1993; Theunissen et al., 2001)]. For simplicity in this paper we use the term *mapping function* instead. Although mapping function estimation has been applied to many sensory modalities (Arabzadeh et al., 2005; DiCaprio, 2003; DiCarlo et al., 1998) this paper focuses on vision and audition.

Many algorithms have been developed to estimate a neuron's mapping function from an ensemble of stimulus-response pairs. These include the spike-triggered average, normalized reverse correlation, linearized reverse correlation, ridge regression, other forms of regularized linear regression, local spectral reverse correlation, spike-triggered covariance, artificial neural networks, maximally informative dimensions, kernel regression, boosting, and models based on leaky integrate-and-fire neurons. Because many of these system identification algorithms were developed in or borrowed from other disciplines, they seem very different superficially and bear little obvious relationship with each other. Each algorithm makes different assumptions about the neuron and how the data is generated. Without a unified framework it is difficult to select the most suitable algorithm for estimating the mapping function. In this review, we present a unified framework for describing these algorithms called *maximum a posteriori* (MAP) estimation (see (Robert, 2001) for a detailed description.

In the MAP framework, the implicit assumptions built into any system identification algorithm are made explicit in three MAP constituents: *model class*, *noise distributions*, and *priors*. Understanding the interplay between these three MAP constituents will simplify the task of selecting the most appropriate algorithms for a given data set. The MAP framework can also facilitate the development of novel system identification algorithms by incorporating biophysically plausible assumptions and mechanisms into the MAP constituents.

In this review we will first introduce the MAP framework and define the three MAP constituents (§2). We will demonstrate the effect of different priors in linear models as illustrative examples (§3). Because the three MAP constituents are independent all the priors presented with the linear model are equally applicable to other model classes (§4 and §5). Finally, we will examine the effect of different noise assumptions on the MAP estimate (§6). Throughout this review, we will demonstrate how each existing system identification algorithm can be viewed as a MAP estimate. We will also reveal the underlying assumptions inherent to each algorithm.

**2 The MAP framework**

The task of neuronal system identification can be viewed as an inference problem. Our goal is to infer a functional model that is consistent with the observed stimulus-response data. This model is a stimulus-response mapping function, $y = f(\mathbf{x})$, that maps any stimulus, $\mathbf{x}$, to the response, $y$. In general, $\mathbf{x}$ can be any representation of the stimulus. It may include spatial dimensions, temporal dimensions, or both. In practice $\mathbf{x}$ is generally a high dimensional vector. We refer to the components of $\mathbf{x}$ as *feature channels* or simply *features*. They can be pixels, sound pressures, or any transformation of the raw stimulus. The response, $y$, is usually a scalar representing the response level; it can be an indicator for the presence of spikes, an integral spike count, or a continuous firing rate.

A typical data set for neuronal system identification is a collection of stimulus-response pairs, $\mathcal{D} = \{\mathbf{x}_i, y_i\}_{i=1}^{N}$ that are treated as independent samples. (For data sets where temporal dependence cannot be ignored, see §6.3.) Due to experimental limitations, neurophysiology data are often scarce and noisy. Therefore, reasonable assumptions and prior knowledge about the system can often improve the accuracy of the inference. The MAP framework enables investigators to incorporate prior knowledge into the inference procedure.

To perform any inference about the model based on the data, it is necessary to assume that there exists a joint distribution, $\mathcal{P}$, between the data and the inferred model, $f$. Because the data samples are assumed to be independent, $\mathcal{P}$ can be factored into a product of the joint probability of each sample, $\mathcal{P}(\mathbf{X}, \mathbf{Y}, f) = \prod_{i=1}^{N} p(\mathbf{x}_i, y_i, f)$. The response data is conditionally dependent to the stimulus and the mapping function, hence we can reformulate $p(\mathbf{x}_i, y_i, f) = p(y_i | \mathbf{x}_i, f) p(\mathbf{x}_i, f)$. Furthermore, the stimulus and the model are independent, so we can further factor $p(\mathbf{x}_i, f)$ as $p(\mathbf{x}_i) p(f)$. Hence the joint distribution can be written as

$$\mathcal{P}(\mathbf{X}, \mathbf{Y}, f) = \prod_{i=1}^{N} p(y_i | \mathbf{x}_i, f) p(\mathbf{x}_i) p(f), \quad (0.1)$$

where $\mathbf{X} = [\mathbf{x}_1, \cdots, \mathbf{x}_N]^T$ is the stimulus matrix and $\mathbf{Y} = [y_1, \cdots, y_N]^T$ is the response matrix.

Because we are interested in inferring a model, $f$, the quantity of interest is the posterior distribution of $f$ conditioned on the observed data. It can be shown (Appendix A) that the effective *posterior distribution* is given by

$$\mathcal{P}^*(f \mid \mathbf{X}, \mathbf{Y}) \propto \prod_{i=1}^{N} p(y_i \mid \mathbf{x}_i, f) p(f). \tag{0.2}$$

This posterior gives a probability distribution over all possible $f$. In theory, it is possible to compute the full posterior distribution. This would enable comprehensive Bayesian analysis of a neuron's mapping function, even providing Bayesian confidence intervals on the resulting model. However, it is rare that one could ever collect enough data for such a full Bayesian computation. A tractable alternative is to infer the single most probable model based on the data. This transforms the problem into a simpler one: finding the $f$ that maximizes the posterior distribution. This estimate of $f$ is called the *maximum a posteriori* (MAP) estimate, $f_{MAP}$. By definition, $f_{MAP}$ is the *extremum* (in statistics, it is also referred to as the *mode*) of the posterior distribution.

Equation (0.2) is critical for the understanding of this review. The first term in equation (0.2) is a conditional distribution, $p(y_i \mid \mathbf{x}_i, f)$, called the *effective likelihood*. This is different from the likelihood function, which is defined as the probability of the observed data given the model, $p(y_i, \mathbf{x}_i \mid f)$. However, the likelihood can always be factored as

$$p(y_i, \mathbf{x}_i \mid f) = p(y_i \mid \mathbf{x}_i, f) p(\mathbf{x}_i \mid f)$$

The model and the stimulus are independent, so the second factor of the likelihood function is $p(\mathbf{x}_i \mid f) = p(\mathbf{x}_i)$. Because $p(\mathbf{x}_i)$ is independent of $f$, it has no effect on the MAP estimate.

Therefore, the effective likelihood (the part of the likelihood that depends on $f$) is $p(y_i | \mathbf{x}_i, f)$. However, the effective likelihood is the only likelihood we will consider in this review, so we here will refer to $p(y_i | \mathbf{x}_i, f)$ as the *likelihood*. This likelihood gives the relative probability for models that could have generated the observed data. The model that maximizes the likelihood is therefore the *maximum likelihood* (ML) estimate, $f_{ML}$. The ML estimate is unbiased in the limit of infinite data, and $f_{ML}$ is the model that gives the best fit to the observed data regardless of any prior knowledge or subjective assumptions about the system under study.

The second term in equation (0.2) is the *prior distribution*, $p(f)$. This prior describes the probability of a model. It is based on subjective assumptions and expert knowledge that are independent of the current data set. Unlike the ML estimate, the MAP estimate includes a subjective prior, and this prior will tend to bias $f_{ML}$ toward models that are believed to be more plausible. However, the posterior distribution will generally grow sharper as the sample size grows, hence the effect of this subjective prior diminishes as more data become available. For a sharply peaked posterior distribution the mode, $f_{MAP}$, will approach the mean, so, the MAP estimate is asymptotically unbiased.

*2.1 Constituents of the MAP estimator*

To compute the effective posterior distribution, the two terms of equation (0.2) -- the likelihood, $p(y_i | \mathbf{x}_i, f)$, and the prior, $p(f)$ -- must be defined. However, these two probability distributions depend on the mapping function, $f$, which is unknown. Thus, there are actually three unknown quantities in equation (0.2): (1) the mapping function, (2) the likelihood and (3)

the prior. These three quantities are defined by the three constituents of the MAP framework: (1) the model class, (2) the noise distribution and (3) the prior. Because these MAP constituents are required before any inference can be made they must be assumed.

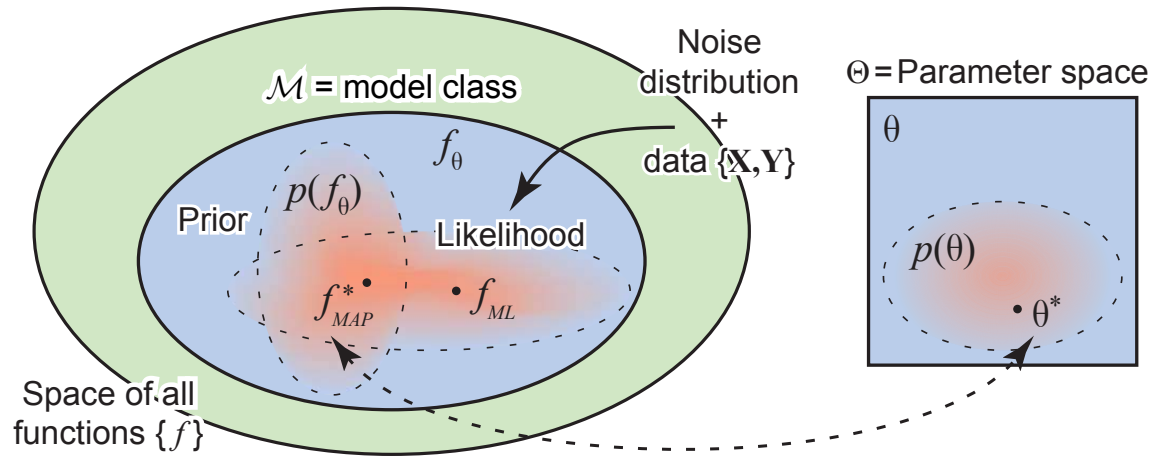

**Figure 1: Maximum a posteriori (MAP) Inference.** The task of an inference algorithm is to find a function, $f$, in a function space, such that $f$ fits the data, $\{\mathbf{X},\mathbf{Y}\}$, well. Typical function spaces (green ellipse) are too large for this purpose, and they contain many functions that can fit the data perfectly. MAP inference is a way to restrict the size of the inferential space through assumptions on the three MAP constituents, so that a unique function, $f^*$, may be obtained. The three MAP constituents are: (1) the model class, (2) the noise distribution, and (3) the prior. The model class, $\mathcal{M}$, defines a parameterized function class (blue ellipse) over which inference is performed. This strongly restricts the inferential space, because $\mathcal{M}$ contains only functions, $f_\theta$, parameterized by the parameter vector, $\boldsymbol{\theta}$. The model class also establishes a one-to-one correspondence between $\boldsymbol{\theta}$ and $f_\theta$, so that inference may be performed over the parameter space, $\Theta$, (a vector space) rather than $\mathcal{M}$ (a function space). The noise distribution together with the data defines a probability distribution, the likelihood, over $\mathcal{M}$. The mode of the likelihood is a function, $f_{ML}$, that fits the data best under the assumed noise distribution. The prior defines a subjective distribution over $\mathcal{M}$ (or equivalently over $\Theta$), that specify the plausibility of the functions (or parameters values) in $\mathcal{M}$ (or in $\Theta$). Multiplying the prior by the likelihood gives the posterior distribution. The mode of the posterior distribution is the MAP estimate, $f^*_{MAP}$, that not only fits the data well, but is also plausible under the assumed prior.

The first MAP constituent is the *model class*. It defines a space of functions over which inference will occur. The search for the optimal mapping function will be constrained within this inferential space. A good model class should contain functions that can provide a good description of the neuron, but are also sufficiently restricted so they do not overfit to the noise in the data. This can be achieved by restricting the model class to be a parameterized class (Figure 1). Functions in such model class are indexed by a set of model parameters that are stored in a parameter vector, $\boldsymbol{\theta}$. Choosing a model class determines the form of the mapping function and its parameters. This, however, imposes strong constraints on the inference problem, as it limits the description of the neuron to the available functions within the assumed model class. In practice, most system identification studies work with a parameterized model class where $f$ is indexed by $\boldsymbol{\theta}$. Therefore, we will write $f_{\boldsymbol{\theta}}$ instead of $f$. (The main distinction between parametric and nonparametric models is that the number of model parameters in nonparametric models can grow with the amount of data available. However, both parametric and nonparametric models are parameterized model classes. For further clarification, see §5). Examples of different model classes used in neuronal system identification are presented in §3, §4 and §5.

The second MAP constituent is the *noise distribution*. It describes stochastic variability in the response that is not explained by the model. Because neurons are stochastic, the response of any neuron to repeated presentation of the stimulus might differ across trials (Geisler and Goldberg, 1966). Although this inherent randomness cannot be predicted, it can be characterized probabilistically. The noise distribution is the probability distribution of the response conditioned on the predicted mean: $p(y_i | \mu_i)$, where $\mu_i = \mathbb{E}(y_i) \propto f_{\boldsymbol{\theta}}(\mathbf{x}_i)$. Choosing a noise distribution will determine which part of the neuronal response variability is predictable versus unpredictable. Thus, any assumption about the noise distribution will also constrain the inference process, because it further constricts the model class to a subset of functions that fit the data well (Figure 1). When the noise distribution is evaluated at the observed data, it is precisely the likelihood in equation (0.2), $p(y_i | \mathbf{x}_i, f_\theta)$. This likelihood assigns a relative probability to each function in the model class based on how well it fits the data. In practice, the most commonly assumed noise distribution is the Gaussian distribution. Other noise distributions used in neuronal system identification are described in §6.

The last MAP constituent is the *prior*. Unlike the model class and the noise distribution, the prior is a subjective probability that describes the plausibility (a subjective belief) of each function in the model class. Because the prior is independent of the data, assumptions and expert knowledge about the system can be incorporated into system identification by choosing an appropriate prior. Selecting a prior will further constrain the inferential space to regions with sufficiently high prior probability density (Figure 1). Because there is a one-to-one correspondence between $f_\theta$ and $\theta$, we will write $p(\theta)$ instead of $p(f_\theta)$ for the prior probability associated with a particular model. Hence equation (0.2) can be written as,

$$\mathcal{P}^*(f | \mathbf{X}, \mathbf{Y}) \propto \prod_{i=1}^{N} p(y_i | \mathbf{x}_i, f) p(f) = \prod_{i=1}^{N} p(y_i | f_\theta(\mathbf{x}_i)) p(\theta)$$

In practice, the most commonly assumed prior is the Gaussian prior. Other examples of priors used in neuronal system identification are presented in §3.

*2.2 Optimization stage of MAP estimation*

Having specified all the unknown quantities of the posterior distribution, $\mathcal{P}^*$, in equation (0.2), it remains to optimize $\mathcal{P}^*$ over all models to obtain the MAP estimate. Direct maximization of $\mathcal{P}^*$ is difficult; instead we maximize the natural log of $\mathcal{P}^*$, or equivalently minimize $-\log \mathcal{P}^*(f \mid \mathbf{X}, \mathbf{Y})$. Due to the correspondence between $f_\theta$ and $\theta$, inferring the most probable model is equivalent to inferring a set of most probable parameters that specify the model. Thus, any inference in the model class is equivalent to an inference in the parameter space. However, it is easier to perform computation in the parameter space (a finite dimensional vector space) than in the model class (an infinite dimensional function space). For example, adding two parameter vectors only requires adding a finite number of components of the parameter space, but adding two functions in the model class would require adding an infinite number of function values. For this reason we will work in the parameter space and minimize $-\log \mathcal{P}^*(\boldsymbol{\theta} \mid \mathbf{X}, \mathbf{Y})$. When minimized over $\boldsymbol{\theta}$, the MAP estimate returns a set of most probable model parameters, $\boldsymbol{\theta}^* = \arg\min_{\boldsymbol{\theta}} \left[ -\log \{\mathcal{P}(\boldsymbol{\theta} \mid \mathbf{X}, \mathbf{Y})\} \right]$, that defines the most probable mapping function, $f_{MAP}$.

$$\boldsymbol{\theta}^* = \arg\min_{\boldsymbol{\theta}} \left[ -\log \left\{ \prod_{i=1}^{N} \overbrace{p(y_i \mid f_\theta(\mathbf{x}_i))}^{\text{Likelihood}} \overbrace{p(\boldsymbol{\theta})}^{\text{Prior}} \right\} \right]$$

(with "Noise", "Model", "Data" labeling the terms inside the likelihood)

$$= \arg\min_{\boldsymbol{\theta}} \left[ -\sum_{i=1}^{N} \log p(y_i \mid f_\theta(\mathbf{x}_i)) - \log p(\boldsymbol{\theta}) \right] \quad (0.3)$$

The function inside the square bracket of equation (0.3) is the *MAP objective function*,

$$g_{MAP}(\boldsymbol{\theta}) = \underbrace{-\sum_{i=1}^{N} \log p(y_i | f_{\boldsymbol{\theta}}(\mathbf{x}_i))}_{\text{Loss Functional}} \underbrace{- \log p(\boldsymbol{\theta})}_{\text{Regularizer}}. \qquad (0.4)$$

The MAP objective is the function to be minimized in order to obtain the MAP estimate, and it consists of two terms. The first term is the *loss functional* and the second term is the *regularizer*. As opposed to the inferential quantities, such as the noise distribution and the prior, the loss functional and the regularizer are quantities we optimize. The intimate relationship and the subtle difference between the inferential quantities and the optimized quantities are presented below.

2.2.1 The loss functional and the noise distribution

The *loss functional* is a function that measures the failure of a model to fit the observed stimulus-response data. When a model fails to fit the data, the loss functional will increase its value to penalize the MAP objective. Minimizing the loss functional will thus produce a function that gives the best fit to the observed data. To properly measure the failure of a model, it is necessary to know the noise distribution. Because the noise distribution describes the stochastic variability in the response, failure in predicting this unpredictable variability should not be considered as a failure of the model. Consequently, any model residuals that are likely a result of the unpredictable noise should only be penalized mildly by the loss functional. Conversely, model residuals that are not likely a result of noise should be penalized strongly by the loss functional, since they reflect genuine model failure. The amount of penalty incurred by the loss functional should be a monotonic function of how likely the residual reflects model failure. The proper amount of penalty is defined by the negative log of the noise distribution.

For example, under the commonly assumed Gaussian noise distribution, the remaining variability in the response given the model prediction is a Gaussian. If the mean, $\mu_i$, of a Gaussian distribution is predicted by the model, $\mu_i = f_\theta(\mathbf{x}_i)$, the Gaussian noise distribution can be written as

$$p(y_i | \mu_i = f_\theta(\mathbf{x}_i)) \propto \exp\left(\frac{-[y_i - f_\theta(\mathbf{x}_i)]^2}{2\sigma^2}\right).$$

Here $\sigma^2$ is the noise variance that characterizes the variability in the data. Under the Gaussian noise assumption, most of the noise variance should lie within two standard deviations of $\mu_i$. The appropriate loss functional is then the negative log of a Gaussian,

$$-\log[p(y_i | f_\theta(\mathbf{x}_i))] \propto \frac{[y_i - f_\theta(\mathbf{x}_i)]^2}{2\sigma^2}. \tag{0.5}$$

This is the square loss commonly used in least-squares regression algorithms. Many neuronal system identification algorithms implicitly assume a Gaussian noise distribution by using the square loss. Clearly, the square loss only penalizes small residuals within $2\sigma$ of $\mu_i$, but the penalties will grow rapidly for residuals beyond $2\sigma$ of $\mu_i$.

2.2.2 The regularizer and the prior distribution

Recall that the prior is a probability distribution that describes the plausibility of models base on non-data-dependent criteria. These criteria will impose constraints on the model. Models that satisfy these constraints are considered plausible and so will have a relatively high prior probability. Any constraint violation will make a model implausible by reducing its prior probability. The *regularizer* is a function that measures the amount of constraint violation and which determines how the violations are penalized. Like the loss functional, the regularizer will penalize the MAP objective by increasing its value in proportion to the amount of constraint violation committed by a model. Because the amount of constraint violation depends on the choice of the subjective prior, the regularizer must be defined in terms of the prior. Indeed, the regularizer is defined as the negative log of the prior distribution.

2.2.3 Inference vs. optimization

Thus far we have treated neuronal system identification as a MAP inference problem, but MAP inference includes an optimization step for finding the mode of the posterior distribution. The initial quantities we assume (the MAP constituents) and the final quantity we compute (the MAP estimate) are all probabilistic quantities for inference. However, neuronal system identification can be treated purely as an optimization problem, without any reference to these probabilistic quantities. In fact, many early neuronal system identification algorithms were constructed as an optimization problem without any reference to probabilistic inference. The inference and the optimization perspective of MAP estimation offer two different starting points for analyzing or constructing a neuronal system identification algorithm. Each perspective has its own strengths and weaknesses, and both perspectives will be covered throughout this review.

In the optimization perspective, investigators begin by defining an objective function. An objective function that is easy to optimize allows simple implementation and computation. In addition, the optimal mapping function can be obtained via optimization without reference to the inferential quantities. However, without the inferential quantities, it is difficult to disentangle the hidden assumptions that are jumbled into a single objective function.

In the inference perspective, the investigators start by choosing the three MAP constituents. This forces the investigators to be explicit about their assumptions. For example, by choosing a biophysically realistic noise distribution and a physiologically inspired prior, the inference perspective enables investigators to incorporate realistic assumptions into an estimation algorithm. A drawback of such biophysically motivated MAP constituents, however is that the loss functional and regularizer may be very difficult to minimize, which may result in intractable MAP estimates. Even for cases where we can write down an explicit expression for the MAP objective function, an efficient algorithm for minimizing it may not be available. Thus, the noise distribution and prior for inference should be selected based on the optimization feasibility of the final MAP objective function.

## 3 Linear models and the effects of priors

The linear model is the simplest and the most commonly used model class. The earliest neuronal system identification algorithm, *reverse correlation*, assumes a linear model. In practice, reverse correlation is often implemented by selecting the stimuli that elicit spikes and then computing the average of these spike-triggered stimuli. Therefore, in the neuroscience community reverse correlation is also referred to as the *spike-triggered average* (STA) (Eggermont et al., 1983a; Ringach and Shapley, 2004). The STA was originally developed to study auditory neurons (De Boer, 1967). Later, the STA was used to characterize the mapping function for simple cells in the primary visual cortex [V1, (Jones and Palmer, 1987)].
The STA algorithm can be framed as a MAP estimation problem. This requires specifying the three MAP constituents. The first MAP constituent is the model class. Because the underlying model class for STA is linear, the model class consists of functions of the form $\mathbf{x}^T\boldsymbol{\beta}$. Each mapping function of the linear model class is indexed by a set of coefficients, $\boldsymbol{\beta} = [\beta_1, \beta_2, \cdots, \beta_d]^T$. The model parameters are simply the elements in $\boldsymbol{\beta}$, so $\boldsymbol{\theta} = \boldsymbol{\beta}$. The MAP estimate of the model parameters can be written as

$$\boldsymbol{\beta}^* = \arg\min_{\boldsymbol{\beta}} \left[ -\sum_{i=1}^{N} \log p(y_i | \mathbf{x}_i^T \boldsymbol{\beta}) - \log p(\boldsymbol{\beta}) \right]. \quad (0.6)$$

The second MAP constituent is the noise distribution. Equation (0.5) shows that assuming a Gaussian noise distribution transforms the negative-log-likelihood in equation (0.6) into the square loss. Thus, under the Gaussian assumption the loss functional in equation (0.6) becomes the residual sum of squares as in the familiar least-squares regression:

$$\begin{aligned}\boldsymbol{\beta}^* &= \arg\min_{\boldsymbol{\beta}} \left[ \sum_{i=1}^{N} \frac{(y_i - \mathbf{x}_i^T \boldsymbol{\beta})^2}{2\sigma^2} - \log p(\boldsymbol{\beta}) \right] \\ &= \arg\min_{\boldsymbol{\beta}} \left[ (\mathbf{Y} - \mathbf{X}\boldsymbol{\beta})^2 - 2\sigma^2 \log p(\boldsymbol{\beta}) \right]. \end{aligned} \quad (0.7)$$

The third MAP constituent is the prior, $p(\boldsymbol{\beta})$. As discussed in §2.2, early neuronal system identification algorithms used the square loss for computational convenience. This is equivalent to an important implicit assumption of a Gaussian noise distribution. Hence, in the sections below, we will illustrate the effect of common priors with a linear model under the Gaussian noise assumption. (In theory, these priors may be used with any model class, and any noise distribution.) Because the STA model class is linear and the noise distribution is Gaussian, the loss functional in the examples below will always take the form of residual sum of squares as shown in (0.7). Only the regularizer that depends on the prior will differ.

*3.1 Generalized Gaussian priors*

As explained in section 2.1 the most commonly used prior in neuronal system identification is the Gaussian prior. The most general form of Gaussian prior can be written as a *multivariate Gaussian* (MVG),

$$p(\boldsymbol{\beta}) \sim \mathcal{N}(\boldsymbol{\beta}_0, \mathbf{A}) \propto \exp\left[-\tfrac{1}{2}(\boldsymbol{\beta}-\boldsymbol{\beta}_0)^T \mathbf{A}^{-1}(\boldsymbol{\beta}-\boldsymbol{\beta}_0)\right], \qquad (0.8)$$

where $\boldsymbol{\beta}_0$ is the mean, and $\mathbf{A}$ is the covariance matrix that determines the covariance of the samples drawn from the MVG. (We ignore the normalization constant of the MVG in equation (0.8) because it does not alter the extremum of the posterior distribution, $\mathcal{P}^*$.) In equation (0.8), $\boldsymbol{\beta}_0$ and $\mathbf{A}$ are parameters that define the MVG, and thus specify the prior. To distinguish these from the model parameters, parameters that specify the prior are called *hyperparameters*. Different Gaussian priors can be obtained simply by choosing different $\boldsymbol{\beta}_0$ and $\mathbf{A}$. In nearly all applications, the MVG is centered, so, the mean of the MVG, $\boldsymbol{\beta}_0$, can be set to zero. Hence, without loss of generality, here we only consider Gaussian priors with zero mean,

$$p(\boldsymbol{\beta}) \sim \mathcal{N}(\boldsymbol{\beta}_0 = 0, \mathbf{A}) \propto \exp\left[-\tfrac{1}{2}\boldsymbol{\beta}^T \mathbf{A}^{-1} \boldsymbol{\beta}\right]. \qquad (0.9)$$

Under the prior given by equation (0.9), the MAP estimate for $\boldsymbol{\beta}$ is obtained by the following optimization problem (c.f. equation (0.7)),

$$\boldsymbol{\beta}^* = \arg\min_{\boldsymbol{\beta}} \left[(\mathbf{Y} - \mathbf{X}\boldsymbol{\beta})^2 - 2\sigma^2 \log\left(\exp\left[-\tfrac{1}{2}\boldsymbol{\beta}^T \mathbf{A}^{-1} \boldsymbol{\beta}\right]\right)\right].$$

The MAP objective function that must be minimized is therefore

$$g_{MAP}(\boldsymbol{\beta}) = (\mathbf{Y} - \mathbf{X}\boldsymbol{\beta})^2 + \sigma^2 \boldsymbol{\beta}^T \mathbf{A}^{-1} \boldsymbol{\beta}. \qquad (0.10)$$

Note that the MAP objective (0.10) is a sum of two quadratic functions of $\boldsymbol{\beta}$. Therefore, it has a unique minimum that can be solved analytically by setting the gradient of the MAP objective to zero and then solving for $\boldsymbol{\beta}$:

$$\begin{aligned} 0 &= \nabla_{\boldsymbol{\beta}} \left[ (\mathbf{Y} - \mathbf{X}\boldsymbol{\beta})^2 + \sigma^2 \boldsymbol{\beta}^T \mathbf{A}^{-1} \boldsymbol{\beta} \right] \\ &= -2\mathbf{X}^T (\mathbf{Y} - \mathbf{X}\boldsymbol{\beta}) + 2\sigma^2 \mathbf{A}^{-1} \boldsymbol{\beta} \\ &= (\mathbf{X}^T \mathbf{X} + \sigma^2 \mathbf{A}^{-1}) \boldsymbol{\beta} - \mathbf{X}^T \mathbf{Y} \end{aligned}$$

The MAP estimate for $\boldsymbol{\beta}$ is therefore given by

$$\boldsymbol{\beta}^* = \underbrace{(\mathbf{X}^T \mathbf{X} + \sigma^2 \mathbf{A}^{-1})^{-1}}_{\text{regularized inverse of the feature autocovariance matrix}} \mathbf{X}^T \mathbf{Y}. \tag{0.11}$$

Equation (0.11) is the MAP estimate for a family of models, which we will refer to here as *LGG family* (linear model class, Gaussian noise, and Gaussian prior). However, as we will see in §4 and §5, the LGG solution (0.11) is also applicable to nonlinear models, as long as the model is linear in the parameters. Because the extremum of the posterior distribution is derived analytically, equation (0.11) gives the general solution for the MAP estimate, $\boldsymbol{\beta}^*$, without invoking any optimization algorithm. However, applying the LGG solution will require computing the regularized inverse of the stimulus' feature autocovariance (the term above the curly bracket in equation (0.11)). This step is usually the computational bottleneck of equation (0.11).

*3.2 The flat prior, STA, and normalized reverse correlation*

The simplest Gaussian prior is the *flat prior* (a.k.a. the non-informative prior). This prior is flat (or non-informative), because it does not *a priori* favor any model over another. In other words, the flat prior assumes all models are equally likely, $p(\beta) = c$, for some appropriate normalization constant $c$. Under the flat prior, the regularizer, $-\log p(\beta)$, is a constant independent of $\beta$. This constant regularizer can be ignored because it will not affect the extremum of the posterior. Many early algorithms for neuronal system identification did not include a regularizer, hence they assumed the flat prior implicitly.

To compute the MAP estimate for linear models under the flat prior, it is advantageous to view this prior as a degenerate Gaussian prior with infinite variance, $\mathbf{A} = diag(\infty)$, or equivalently $\mathbf{A}^{-1} = 0$. By setting the inverse of the covariance matrix to zero ($\mathbf{A}^{-1} = 0$) in equation (0.8), the prior, $p(\beta)$, will be an exponential of zero, hence will result in a flat prior. Because a flat prior is really a Gaussian prior with $\mathbf{A}^{-1} = 0$, the MAP estimate can be obtained by substituting $\mathbf{A}^{-1} = 0$ into the LGG equation (0.11)

$$\beta^* = \left(\mathbf{X}^T\mathbf{X} + \sigma^2\mathbf{A}^{-1}\right)^{-1}\mathbf{X}^T\mathbf{Y} = \left(\mathbf{X}^T\mathbf{X}\right)^{-1}\mathbf{X}^T\mathbf{Y}. \qquad (0.12)$$

Equation (0.12) is the solution to the least-squares regression problem. Thus, the MAP estimate under a flat prior gives the same solution as the least-squares regression, which is precisely the ML estimate under Gaussian noise (Dobson, 2002). So the MAP estimate under the flat prior will always reduce to the ML estimate.

In equation (0.12), the term $\mathbf{X}^T\mathbf{Y}$ is merely the response weighted average of the stimulus features, and $(\mathbf{X}^T\mathbf{X})^{-1}$ is the inverse of the stimulus feature autocovariance. Therefore, the MAP estimate of the linear model under the flat prior is precisely the STA, $\mathbf{X}^T\mathbf{Y}$, normalized by the stimulus autocovariance, $\mathbf{X}^T\mathbf{X}$. This algorithm is commonly known as normalized reverse correlation or the normalized STA (David and Gallant, 2005; Woolley et al., 2006).

3.2.1 The effects of stimulus statistics

Using equation (0.12), we can re-derive some of the well-known effects of stimulus statistics on the STA solution. If the stimulus features are white and sampled densely enough so that the stimulus features are uncorrelated, $\mathbf{X}^T\mathbf{X} = \sigma_X \mathbf{I}$, then the inverse of $\mathbf{X}^T\mathbf{X}$ is proportional to the identity matrix. This is also true when the stimuli are generated according to an m-sequence (maximum length sequence). In both cases, the MAP estimate will be proportional to the STA,

$$\boldsymbol{\beta}^* = \sigma_X^{-1} \mathbf{X}^T \mathbf{Y}. \qquad (0.13)$$

Equation (0.13) is the STA used in simple reverse correlation under white noise stimulation (De Boer and Kuyper, 1968). If the stimulus features are white, STA provides an unbiased estimate of the model parameters, $\boldsymbol{\beta}$. However, when the stimulus features are not white, the STA must be normalized by the stimulus autocovariance (Theunissen et al., 2001) as in equation (0.12). Otherwise, the STA will not maximize the posterior, $\mathcal{P}(\boldsymbol{\beta}|\mathbf{X},\mathbf{Y})$, and simple reverse correlation will not yield the most probable model given the data. The STA in (0.13) is derived under the Gaussian noise assumption. However, as long as the stimulus features have a Gaussian white noise distribution, the STA will give an unbiased estimate for $\beta$ under other assumed noise distributions as well (Nykamp and Ringach, 2002).

*3.3 The spherical Gaussian prior and ridge regression*

The simplest non-flat prior is the *spherical Gaussian prior*. This prior assumes the model parameters are all independent from one another, and the uncertainty (*prior variances*) in every parameter is the same. Assuming the spherical Gaussian is centered, $\boldsymbol{\beta}_0 = 0$, this prior's main difference from the MVG is the diagonal covariance matrix $\mathbf{A} = \sigma_{\boldsymbol{\beta}}^2 \mathbf{I}$, where $\sigma_{\boldsymbol{\beta}}^2$ is the prior variance that describe the amount of uncertainty in the prior:

$$p(\boldsymbol{\beta}) \sim \mathcal{N}\left(0, \sigma_{\boldsymbol{\beta}}^2 \mathbf{I}\right)$$
$$\propto \exp\left[-\tfrac{1}{2} \boldsymbol{\beta}^T \left(\sigma_{\boldsymbol{\beta}}^2 \mathbf{I}\right)^{-1} \boldsymbol{\beta}\right]$$
$$= \exp\left[-\tfrac{1}{2\sigma_{\boldsymbol{\beta}}^2} \boldsymbol{\beta}^T \boldsymbol{\beta}\right].$$

Geometrically, this prior assumes that there is a spherical region in the parameter space within which the parameter values are highly plausible (Figure 2a). Substituting the covariance matrix, $\mathbf{A} = \sigma_{\boldsymbol{\beta}}^2 \mathbf{I}$, into equation (0.10), we obtain the MAP objective for linear models under the spherical Gaussian prior:

$$g_{MAP}(\boldsymbol{\beta}) = (\mathbf{Y} - \mathbf{X}\boldsymbol{\beta})^2 + \sigma^2 \boldsymbol{\beta}^T \left(\sigma_{\boldsymbol{\beta}}^{-2} \mathbf{I}\right) \boldsymbol{\beta}$$

$$= (\mathbf{Y} - \mathbf{X}\boldsymbol{\beta})^2 + \lambda \|\boldsymbol{\beta}\|_2^2. \qquad (0.14)$$

Here $\lambda \equiv \sigma^2 / \sigma_{\boldsymbol{\beta}}^2$ is a *regularization hyperparameter* that determines the tradeoff between minimizing the loss functional and the regularizer, and $\|\boldsymbol{\beta}\|_2^2$ is the $\ell_2$-*norm* of the parameter vector, $\boldsymbol{\beta}$. Equation (0.14) is the $\ell_2$-*norm-penalized* least-squares. The MAP estimate is obtained by minimizing the MAP objective (0.14). The minimum is readily obtained by substituting the covariance matrix, $\mathbf{A} = \sigma_{\boldsymbol{\beta}}^2 \mathbf{I}$, into the LGG solution (0.11),

$$\boldsymbol{\beta}^* = \left(\mathbf{X}^T\mathbf{X} + \sigma^2(\sigma_\beta^2\mathbf{I})^{-1}\right)^{-1}\mathbf{X}^T\mathbf{Y}$$

$$= \left(\mathbf{X}^T\mathbf{X} + \lambda\mathbf{I}\right)^{-1}\mathbf{X}^T\mathbf{Y}. \tag{0.15}$$

Equation (0.14) has the same form as the penalized least-squares used in ridge regression (Hoerl and Kennard, 1970). Consequently, equation (0.15) is the ridge regression solution.

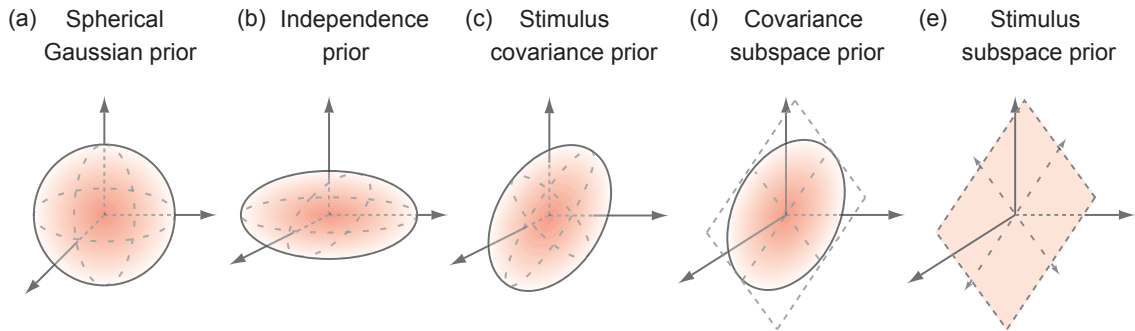

**Figure 2: Gaussian priors.** The regions of high prior probability defined by different Gaussian priors are multidimensional ellipsoids (or their degenerate forms) in the parameter space. (a) The spherical Guassian prior is a special case where the parameters are assumed independent with the same prior variance. Therefore, the ellipsoid degenerates to a sphere, because the principal axes all have the same length, and they are aligned to the axes of the parameter space. (b) The independence prior relaxes the constraint that requires all prior variances to be equal, so the principal axes do not have the same length. But the parameters are still independent, so the principal axes are still axes-aligned. (c) The stimulus covariance prior allows correlations in the parameters so the principal axes of the ellipsoid are not axes-aligned. Instead their direction and length are determined by the eigenvectors and eigenvalues of the stimulus autocovariance matrix. (d) The covariance subspace prior is a degenerate case of the stimulus covariance prior, so the directions of the principal axes are again defined by the eigenvectors of the stimulus autocovariance matrix. However, the directions with small prior variances (with eigenvalues $< \varepsilon$) are collapsed to zero, so the ellipsoid lies on a lower dimensional hyperplane. (e) The stimulus subspace prior is another degenerate case of the stimulus covariance prior, so it lies in the same subspace as the covariance subspace prior. However, the principal axes within the subspace have infinite length, so the lower dimensional ellipsoid degenerate into a lower dimensional hyperplane.

### 3.3.1 Relationship between priors and regularization

Assuming a prior distribution always leads to a form of regularization, but the relationship between prior and regularization is often obscured. *Regularization* is any procedure that restricts the complexity of a model to prevent overfitting. In the MAP framework, regularization is achieved by adding penalty terms to the MAP objective function. Because these penalty terms are the regularizers introduced by the prior distribution, there is also an intimate relationship between the prior and regularization. Using the spherical Gaussian prior as an example, we will clarify the relationship between the prior distribution and regularization.

The regularization hyperparameter, $\lambda \equiv \sigma^2/\sigma_{\boldsymbol{\beta}}^2$, in equation (0.14) is a free parameter that determines the relative importance between minimizing the loss functional, $(\mathbf{Y} - \mathbf{X}\boldsymbol{\beta})^2$, and the regularizer, $\|\boldsymbol{\beta}\|_2^2$. Consequently, $\lambda$ depends on the uncertainties in each respective term. The uncertainty in the loss functional indicates how noisy, hence, uncertain our data are. This uncertainty is characterized by the noise variance, $\sigma^2$. The uncertainty in the regularizer indicates how doubtful we are about the assumed prior. This uncertainty is defined by the prior variance, $\sigma_{\boldsymbol{\beta}}^2$. For example, when we are unsure about the prior we assume, we should choose a relatively flat and broad prior that will not strongly favor one model over another. Computationally, when we are unsure about the assumed prior, $\sigma_{\boldsymbol{\beta}}^2$ will be large compared to $\sigma^2$ and $\lambda \equiv \sigma^2/\sigma_{\boldsymbol{\beta}}^2$ will be small. Because the regularizer in equation (0.14) is multiplied by a small $\lambda$, the value of the objective function will be dominated by the loss functional. Therefore, when the prior is uncertain, the best model fit will be determined by minimizing the loss functional. On the other hand, when the data are noisy, $\sigma^2$ will be large compared to $\sigma_{\boldsymbol{\beta}}^2$, so $\lambda \equiv \sigma^2/\sigma_{\boldsymbol{\beta}}^2$ will be large. Because the regularizer in equation (0.14) is multiplied by a large $\lambda$, the value of the objective function will be dominated by the regularizer. Therefore, when the data are uncertain, the best model fit will be determined by minimizing the regularizer, i.e., reducing the $\ell_2$-norm of the model parameters. This prevents overfitting the data, and thus serves as a means of regularization.

To understand how a prior governs the tradeoff between model accuracy and model complexity, we will take a geometric perspective. The prior variance, $\sigma_\beta^2$, determines the radius, hence the volume, of the Gaussian sphere in the parameter space (Figure 2a). A small $\sigma_\beta^2$, indicating that we are very certain about our prior, will significantly restrict the size of the model class (Figure 1). Consequently, few potential mapping functions will be available to fit the data. This will decrease model complexity and reduce model accuracy. A large $\sigma_\beta^2$, indicating that we are uncertain about our prior, will mildly restrict the size of the model class. Hence, the set of potential mapping functions will still be quite large. This will offer many functions to fit the data. As a result, model accuracy will increase, but so will the model complexity.

3.3.2 Determining the value of $\lambda$

Recall that the regularization hyperparameter, $\lambda$, is defined by two quantities: the noise variance, $\sigma^2$, and the prior variance, $\sigma_\beta^2$. In practice, only the relative values between $\sigma^2$ and $\sigma_\beta^2$ are relevant, hence only the value of $\lambda$ needs to be determined. There are several ways to determine the proper values of $\lambda$ (Bengio, 2000; Golub et al., 1979; Mackay, 1995; Stone, 1974). Because there is only one hyperparameter, *cross-validation* offers a simple and objective way to determine the optimal value of $\lambda$ (Bengio, 2000; Stone, 1974). However, as the number of hyperparameter increases, more efficient methods for hyperparameter optimization, such as *automatic relevancy determination* [ARD, (§3.4.1)], random search (Bergstra and Bengio, 2012) or Bayesian methods (Snoek et al., 2012) may be necessary.

*3.4 The independence prior and automatic relevancy determination*

The *independence prior* can be viewed as a generalization of the spherical Gaussian prior. Recall that the prior variances for the spherical Gaussian prior are equal for every component of the parameter vector ($\sigma^2_{\beta_j} = \sigma^2_{\boldsymbol{\beta}}$ for all $j = 1, 2, \cdots, d$). The independence prior relaxes this constraint, so the prior variances, $\sigma^2_{\beta_j}$, for each component, $\beta_j$, of $\boldsymbol{\beta}$ may be different. Because the independent Gaussian variables are necessarily uncorrelated, the covariance matrix among the components of $\boldsymbol{\beta}$ is still diagonal. Therefore, the independence prior is characterized by the diagonal covariance matrix $\mathbf{A} = \sigma^2_{\beta_j} \mathbf{I} = diag\left(\sigma^2_{\beta_j}\right)$, where $j = 1, 2, \cdots, d$. The independence prior can be written as:

$$\begin{aligned}
p(\boldsymbol{\beta}) &\sim \mathcal{N}\left(0, diag\left(\sigma^2_{\beta_j}\right)\right) \\
&\propto \exp\left[-\tfrac{1}{2}\boldsymbol{\beta}^T \left[diag\left(\sigma^2_{\beta_j}\right)\right]^{-1} \boldsymbol{\beta}\right] \\
&= \prod_{j=1}^{d} \exp\left[-\frac{1}{2}\sigma^{-2}_{\beta_j}\beta^2_j\right].
\end{aligned} \quad (0.16)$$

Geometrically, the region of high prior probability defined by the independence prior is a Gaussian ellipsoid in the parameter space (Figure 2b). The axes of this ellipsoid are aligned to the axes of the parameter space and their lengths are defined by the prior variances, $\sigma^2_{\beta_j}$. Using the independence prior (0.16), the MAP objective function (0.10) can be written as:

$$g_{MAP}(\boldsymbol{\beta}) = (\mathbf{Y} - \mathbf{X}\boldsymbol{\beta})^2 + \sigma^2 \boldsymbol{\beta}^T \left[diag\left(\sigma^{-2}_{\beta_j}\right)\right]\boldsymbol{\beta}$$

$$= (\mathbf{Y} - \mathbf{X}\boldsymbol{\beta})^2 + \sum_{j=1}^{d} \lambda_j \beta_j^2 \,. \tag{0.17}$$

Here $\lambda_j \equiv \sigma^2 / \sigma_{\beta_j}^2$ are the regularization hyperparameters. Since the prior variances are different for each model parameter $\beta_j$, there is one regularization hyperparameter for each $\beta_j$. Because the independence prior is also a Gaussian prior with covariance matrix $\mathbf{A} = diag(\sigma_{\beta_j}^2)$, the minimum of the MAP objective (0.17) is readily obtained by the LGG solution (0.11),

$$\boldsymbol{\beta}^* = \left(\mathbf{X}^T \mathbf{X} + \sigma^2 diag(\sigma_{\beta_j}^{-2})\right)^{-1} \mathbf{X}^T \mathbf{Y} \,.$$

3.4.1 Automatic relevancy determination

In contrast to the spherical Gaussian prior, the MAP objective (0.17) of the independence prior (0.16) has many regularization hyperparameters, $\lambda_j$, for $j = 1, 2, \cdots, d$. Although the values of $\lambda_j$ may be determined by cross-validation, this method must re-minimize the MAP objective for each combination of $\lambda_j$, so the computational load for cross-validation grows exponentially with the number of hyperparameters. A more efficient way to simultaneously determine the value of many $\lambda_j$ is *evidence maximization* (Bengio, 2000; Mackay, 1995). The *evidence* is defined as the expectation of the likelihood with respect to the prior distribution over the model parameters:

$$p(\mathbf{Y} | \mathbf{X}, \sigma, \sigma_{\beta_j}^2) \equiv \int p(\mathbf{Y} | \mathbf{X}, \boldsymbol{\beta}, \sigma) p(\boldsymbol{\beta} | \sigma_{\beta_j}^2) d\boldsymbol{\beta} \,. \tag{0.18}$$

Because the prior variances, $\sigma^2_{\beta_j}$, are the parameters for the prior, $p(\boldsymbol{\beta}|\sigma^2_{\beta_j})$, and the noise variance, $\sigma$, is also a parameter of the likelihood, $p(\mathbf{Y}|\mathbf{X},\boldsymbol{\beta},\sigma)$, the evidence is a function of $\sigma^2_{\beta_j}$ and $\sigma$. The values of $\sigma^2_{\beta_j}$ and $\sigma$ are chosen so that the evidence (0.18) is maximized. In contrast to cross validation, evidence maximization sets the value of $\lambda_j$ by optimizing the value of $\sigma^2$ and $\sigma^2_{\beta_j}$ simultaneously.

One popular implementation of evidence maximization for regression models is *automatic relevancy determination* (ARD, Mackay, 1992, 1995). ARD approximates the evidence by a MVG distribution. This enables the use of gradient-based algorithms or fix-point iterations to maximize the evidence efficiently(Sahani and Linden, 2003). ARD is usually employed for *feature selection*: a procedure that chooses the relevant regression variables for determining the model output. For linear models, there is exactly one model parameter, $\beta_j$, for each input feature, so the independence prior yields a MAP objective with one regularization hyperparameter, $\lambda_j$, for each $\beta_j$. During the minimization of the MAP objective, each $\lambda_j$ can decrease the value of $\beta_j$ independently in order to reduce the effect of the $j$th input feature. An extremely large value of $\lambda_j$ can even drive $\beta_j$ to zero, removing the effect of the $j$th input feature entirely, and therefore achieving automatic feature selection.

*3.5 The correlated Gaussian prior and the stimulus covariance prior*

The Gaussian priors discussed so far assume no correlations among the model parameters, $\beta_j$. Therefore the covariance matrices of these Gaussian priors are diagonal, and the ellipsoids of high prior probability are aligned to the axes of the parameter space. However, when non-white stimuli are used, it may be more appropriate to assume some correlations among the various $\beta_j$. This assumption can be imposed by using *correlated Gaussian priors* with non-diagonal covariance matrix, $\mathbf{A}$. These priors are easier to work with in the eigenspace where the input features are decorrelated and the feature autocovariance matrix, $\mathbf{X}^T\mathbf{X}$, is diagonalized. Because linear transformation of a Gaussian produces another Gaussian, a Gaussian prior in the eigenspace will remain a Gaussian prior. Geometrically, a correlated Gaussian prior can be viewed as a Gaussian ellipsoid that is not aligned to the axes of the parameter space (Figure 2c). For natural stimuli that are highly correlated, the feature autocovariance matrix, $\mathbf{X}^T\mathbf{X}$, is not diagonal. However, because $\mathbf{X}^T\mathbf{X}$ is a covariance matrix and the number of features are less than the number of data points ($d < N$), it is positive definite. Consequently, it can be diagonalized by an orthogonal matrix, $\mathbf{Q}$, so that

$$\mathbf{X}^T\mathbf{X} = \mathbf{Q}\mathbf{D}\mathbf{Q}^T. \qquad (0.19)$$

Here $\mathbf{D} = diag(\delta_j)$ is diagonal matrix, with $\delta_j$ the eigenvalues of $\mathbf{X}^T\mathbf{X}$. Linear transformation by $\mathbf{Q}$ will rotate the coordinate into an eigenspace where $\mathbf{X}^T\mathbf{X}$ is diagonal. Therefore, specifying a prior in the eigenspace amounts to choosing a covariance matrix of the form $\mathbf{A} = \mathbf{Q}\mathbf{C}\mathbf{Q}^T$. Because $\mathbf{Q}$ is fixed by the data, only $\mathbf{C}$ needs to be specified. For example, if $\mathbf{C} = diag(\sigma_{\beta_j}^2)$ then the resulting Gaussian prior is an independence prior in the eigenspace. If $\mathbf{C} = \mathbf{D}$, where $\mathbf{D}$ is as defined in equation (0.19), then the resulting prior is the *stimulus covariance prior*. The stimulus covariance prior is a Gaussian prior with the same covariance structure as the stimulus, so that $\mathbf{A} = \mathbf{Q}\mathbf{D}\mathbf{Q}^T = \mathbf{X}^T\mathbf{X}$. Using this prior means that the model parameters, $\beta_j$, are assumed to be correlated with the same covariance structure as the stimulus. The stimulus covariance prior and the independence prior are similar in that the Gaussian ellipsoids defined by both priors are identical up to a rotation and scaling of the axes (Figure 2b and 2c). This similarity is easily understood when we define $\boldsymbol{\beta}' = \mathbf{Q}^T\boldsymbol{\beta}$ as the model parameters in the eigenspace. Making use of the orthogonality of $\mathbf{Q}$ and given that the eigenvalues are strictly positive, the MAP objective in the eigenspace can then be written as:

$$\begin{aligned} g_{MAP}(\boldsymbol{\beta}') &= (\mathbf{Y} - \mathbf{X}\boldsymbol{\beta})^2 + \sigma^2 \boldsymbol{\beta}^T \left[\mathbf{Q}\mathbf{D}\mathbf{Q}^T\right]^{-1} \boldsymbol{\beta} \\ &= (\mathbf{Y} - \mathbf{X}\mathbf{Q}\mathbf{Q}^T\boldsymbol{\beta})^2 + \sigma^2 \boldsymbol{\beta}^T \left[\underbrace{\left(\mathbf{Q}^T\right)^{-1}}_{\mathbf{Q}^{-1}} \mathbf{D}^{-1} \underbrace{\mathbf{Q}^{-1}}_{\mathbf{Q}^T}\right] \boldsymbol{\beta} \\ &= (\mathbf{Y} - \mathbf{X}\mathbf{Q}\boldsymbol{\beta}')^2 + \sigma^2 \boldsymbol{\beta}^T \left[\mathbf{Q}\mathbf{D}^{-1}\mathbf{Q}^T\right] \boldsymbol{\beta} \\ &= (\mathbf{Y} - \mathbf{X}\mathbf{Q}\boldsymbol{\beta}')^2 + \sigma^2 \left(\mathbf{Q}^T\boldsymbol{\beta}\right)^T \mathbf{D}^{-1} \left(\mathbf{Q}^T\boldsymbol{\beta}\right) \\ &= (\mathbf{Y} - \mathbf{X}\mathbf{Q}\boldsymbol{\beta}')^2 + \sigma^2 \boldsymbol{\beta}'^T \left[diag\left(\delta_j^{-1}\right)\right]\boldsymbol{\beta}' \\ &= (\mathbf{Y} - \mathbf{X}\mathbf{Q}\boldsymbol{\beta}')^2 + \sum_{j=1}^d \lambda_j \beta_j'^2 . \end{aligned} \qquad (0.20)$$

Here $\lambda_j = \sigma^2/\delta_j$, where $\delta_j$ take the roles of the prior variances in the eigenspace. The striking similarity between equation (0.20) and equation (0.17) is apparent. However, the independence prior, the prior variances, $\delta_j$, are fixed by the choice of the stimulus. The only free parameters of the stimulus covariance prior is the noise variance, $\sigma^2$, which can be determined by cross validation.

Minimizing the MAP objective (0.20) gives the MAP estimate of the parameter in the eigenspace. The result can be transformed back to the original parameter space via $\boldsymbol{\beta} = \mathbf{Q}\boldsymbol{\beta}'$.

Making use of the orthogonality of $\mathbf{Q}$, the MAP estimate of $\boldsymbol{\beta}^*$ under the stimulus covariance prior can also be obtained readily from the LGG solution (0.11),

$$\begin{aligned}
\boldsymbol{\beta}^* &= \left[\mathbf{X}^T\mathbf{X} + \sigma^2 \mathbf{A}^{-1}\right]^{-1} \mathbf{X}^T \mathbf{Y} \\
&= \left[\underbrace{\mathbf{X}^T\mathbf{X}}_{\mathbf{QDQ}^T} + \sigma^2 \left(\underbrace{\mathbf{X}^T\mathbf{X}}_{\mathbf{QDQ}^T}\right)^{-1}\right]^{-1} \mathbf{X}^T \mathbf{Y} \\
&= \left[\mathbf{QDQ}^T + \sigma^2 \left(\mathbf{Q}^T\right)^{-1} \mathbf{D}^{-1} \mathbf{Q}^{-1}\right]^{-1} \mathbf{X}^T \mathbf{Y} \\
&= \left[\mathbf{QDQ}^T + \sigma^2 \mathbf{Q}\mathbf{D}^{-1}\mathbf{Q}^T\right]^{-1} \mathbf{X}^T \mathbf{Y} \\
&= \left[\mathbf{Q}\left(\mathbf{D} + \sigma^2 \mathbf{D}^{-1}\right)\mathbf{Q}^T\right]^{-1} \mathbf{X}^T \mathbf{Y} \\
&= \left(\mathbf{Q}^T\right)^{-1} \left(\mathbf{D} + \sigma^2 \mathbf{D}^{-1}\right)^{-1} \mathbf{Q}^{-1} \mathbf{X}^T \mathbf{Y} \\
&= \mathbf{Q}\left[\operatorname{diag}\left(\frac{\delta_j^2 + \sigma^2}{\delta_j}\right)\right]^{-1} \mathbf{Q}^T \mathbf{X}^T \mathbf{Y} \\
&= \mathbf{Q}\left[\operatorname{diag}\left(\frac{\delta_j}{\delta_j^2 + \sigma^2}\right)\right] \mathbf{Q}^T \mathbf{X}^T \mathbf{Y}.
\end{aligned} \quad (0.21)$$

3.5.1 Wiener filter and shrinkage filter

The stimulus covariance prior reveals an interesting link between MAP estimation and Wiener filtering. To see this link, consider a Gaussian signal with variance $\sigma_s^2$ corrupted with additive Gaussian noise with variance $\sigma_n^2$. The optimal filter for recovering the signal is given by the *Wiener filter*, $\sigma_s^2/(\sigma_s^2 + \sigma_n^2)$ (Portilla et al., 2003). (The Wiener filter is a linear filter that will pass on strong signals with $\sigma_s^2 \gg \sigma_n^2$, and will suppress weak signals with $\sigma_s^2 \ll \sigma_n^2$ by a factor of $\sim 1/\sigma_n^2$.) In addition, consider the following equivalence relationship, which is obtained from $\mathbf{X}^T\mathbf{X} = \mathbf{QDQ}^T$ [equation (0.19)] via pre-multiplying by $\mathbf{Q}^T$ and post-multiplying by $\mathbf{X}^{-1}$:

$$\mathbf{Q}^T(\mathbf{X}^T\mathbf{X})\mathbf{X}^{-1} = \mathbf{Q}^T(\mathbf{Q}\mathbf{D}\mathbf{Q}^T)\mathbf{X}^{-1}$$
$$\mathbf{Q}^T\mathbf{X}^T = \mathbf{D}\mathbf{Q}^T\mathbf{X}^{-1}.$$

Here $\mathbf{X}^{-1}$ is only a left inverse and may not even be a square matrix. Substituting $\mathbf{D}\mathbf{Q}^T\mathbf{X}^{-1}$ for $\mathbf{Q}^T\mathbf{X}^T$ in equation (0.21), the MAP estimate can be re-expressed as

$$\boldsymbol{\beta}^* = \mathbf{Q}\left[\underbrace{diag\left(\frac{\delta_j^2}{\delta_j^2+\sigma^2}\right)}_{\text{Wiener Filter}}\right]\mathbf{Q}^T\mathbf{X}^{-1}\mathbf{Y}, \quad (0.22)$$

where $\delta_j$ are the eigenvalues of $\mathbf{X}^T\mathbf{X}$. Because $\mathbf{X}^{-1}$ is usually ill-conditioned and singular, we never actually compute the MAP estimate with equation (0.22). However, it is clear from equation (0.22) that the MAP estimate under the stimulus covariance prior takes the form of a Wiener filter. This Wiener filter passes information from channels with $\delta_j^2 \gg \sigma^2$, and it suppresses information from channels with $\delta_j^2 \ll \sigma^2$ by a factor of $\sim 1/\sigma^2$. Because $\sigma^2 > 0$, the Wiener filter will always be less than unity. Therefore, using the stimulus covariance prior also has the effect of applying a *shrinkage filter*.

*3.6 The subspace priors and the regularized pseudo-inverse*

The eigenvalues, $\delta_j$, for the stimulus autocovariance, $\mathbf{X}^T\mathbf{X}$, are determined by the nature of the stimulus. If the stimulus is white noise, then all $\delta_j$ should be roughly the same, resulting in a white spectrum. On the other hand, if a natural stimulus is used then the spectrum of $\delta_j$ decays according to an inverse power law (Field, 1987). This means that the high-frequency features in natural stimuli will be much more scarce than the low-frequency features.

Consequently, the model parameters for the low-frequency features will have high *signal to noise ratio* (SNR), and these parameters will be relatively less susceptible to noise contamination.

Because natural stimuli constitute a finite and sparse sampling of the stimulus space, the low-SNR model parameters usually cannot be estimated accurately. To incorporate this prior knowledge into an estimation algorithm, the *subspace priors* can be used. The primary effect of these priors is noise thresholding. The subspace priors have one free parameter, $\varepsilon$, representing a noise threshold. Any parameters with SNR lower than $\varepsilon$ are likely to be masked by noise. These priors will then eliminate these noisy parameters by setting them to zero, and eliminating the effects of the noisy high-frequency features. Here we present two examples of subspace priors: the covariance subspace prior (§3.6.1) and the stimulus subspace prior (§3.6.2).

3.6.1 The covariance subspace prior

The *covariance subspace prior* eliminates the low-SNR parameters with a noise threshold, $\varepsilon$, and assumes a stimulus covariance prior over the high-SNR parameters. This prior is specified by the covariance matrix $\mathbf{A} = \mathbf{Q}\mathbf{D}_\varepsilon\mathbf{Q}^T$, where $\mathbf{Q}$ is the usual orthogonal matrix containing the eigenvectors of $\mathbf{X}^T\mathbf{X}$. And the diagonal matrix, $\mathbf{D}_\varepsilon$, contains the eigenvalue, $\delta_j$, of $\mathbf{X}^T\mathbf{X}$ with the small eigenvalues, $\delta_j \leq \varepsilon$, set to zero,

$$\mathbf{D}_\varepsilon = diag(\delta_j > \varepsilon). \qquad (0.23)$$

This thresholding procedure is similar to a well-known dimensionality reduction algorithm, known as *principal component analysis* (PCA). Geometrically, the covariance subspace prior can be viewed as a degenerate stimulus covariance prior whose high dimensional Gaussian ellipsoid (Figure 2c) is collapsed down to a lower dimensional ellipsoid (Figure 2d). This effectively reduces the dimensionality of the Gaussian ellipsoid in the parameter space.

Setting the eigenvalues, $\delta_j$, to zero has dramatic effects on the interpretation and computation of the covariance subspace prior. Because $\delta_j$ represent the prior variances in the eigenspace, they characterize the uncertainty of the prior about the model parameters, $\beta'_j$. [Recall that $\beta'_j$ is the parameter in the eigenspace (§3.5).] Setting a particular $\delta_j = 0$ implies that the prior has zero uncertainty and therefore the values of $\beta'_j$ are known. Because a Gaussian distribution with zero variance degenerates to a *delta-function*, the covariance subspace prior can be written as a product of Gaussian distributions and delta-functions. Moreover, a delta-function will fix the value of a particular $\beta'_j$, so that the region of high prior probability along the $\beta'_j$ axis will collapsed to a point. The essence of the covariance subspace priors is to collapse the region of high prior probability in the parameter space down to a lower dimensional subspace. This significantly reduces the size of the model class. Assuming the stimulus covariance prior over the high-SNR parameters further restricts the parameter values within the subspace to a Gaussian ellipsoid (Figure 2d).

The MAP objective for a linear model under Gaussian noise and the covariance subspace prior is

$$g_{MAP}(\boldsymbol{\beta}') = (\mathbf{Y} - \mathbf{XQ}\boldsymbol{\beta}')^2 + \sum_{j, \delta_j > \varepsilon} \lambda_j \beta'^2_j + \sum_{j, \delta_j \leq \varepsilon} \infty \cdot \beta'^2_j. \qquad (0.24)$$

Note that when $\beta'_j \neq 0$ for any $\delta_j \leq \varepsilon$, the value of the MAP objective will be penalized infinitely by the last term of equation (0.24). Because we want to minimize the MAP objective, an infinite penalty ensures that $\beta'_j \neq 0$ will never happen for any $j$ with $\delta_j \leq \varepsilon$. The minima of equation (0.24) can be obtained from the LGG solution (0.11) as well. However, by setting some prior variances to zero, the regularized inverse in equation (0.11) becomes singular. Hence the matrix inversion in equation (0.11) must be replaced by a pseudo-inverse that is usually computed via singular value decomposition. For this reason, the regularized inverse in (0.11) is also called the *regularized pseudo-inverse*. The MAP estimate for the parameter can be written as

$$\boldsymbol{\beta}^* = \left[\mathbf{Q}\mathbf{D}_\varepsilon\mathbf{Q}^T\right]^+ \mathbf{X}^T\mathbf{Y}, \qquad (0.25)$$

where $\mathbf{D}_\varepsilon$ is given by equation (0.23), and the superscript-plus denotes the pseudo-inverse. Finally, because the covariance subspace prior has only one regularization hyperparameter, the noise threshold, $\varepsilon$, its value can be determined by cross-validation (David et al., 2004; Theunissen et al., 2001; Woolley et al., 2005).

### 3.6.2 The stimulus subspace prior

The *stimulus subspace prior* also eliminates the low-SNR parameters that tend to amplify noise, but it assumes a flat prior over the high-SNR parameters. This stimulus subspace prior is defined by the covariance matrix $\mathbf{A} = \mathbf{Q}\mathbf{P}_\varepsilon\mathbf{Q}^T$, where $\mathbf{Q}$ is the usual eigenvector matrix of $\mathbf{X}^T\mathbf{X}$. And $\mathbf{P}_\varepsilon$ is derived from the eigenvalue matrix, $\mathbf{D} = diag(\delta_j)$, of $\mathbf{X}^T\mathbf{X}$ as follow:

$$\left[\mathbf{P}_\varepsilon\right]_{ii} = \begin{cases} 0, & [\mathbf{D}]_{ii} \leq \varepsilon \\ \infty, & [\mathbf{D}]_{ii} > \varepsilon \end{cases} \qquad (0.26)$$

Thus, $\mathbf{P}_\varepsilon$ is obtained by replacing the small eigenvalues ($\delta_j \leq \varepsilon$) of $\mathbf{D}$ with zeros, and replacing the large eigenvalues ($\delta_j > \varepsilon$) by infinities as in the flat prior (§3.2). Geometrically, the stimulus subspace prior can be viewed as a combination of two degenerate Gaussian priors over two different groups of model parameters in the eigenspace of $\mathbf{X}^T\mathbf{X}$. Thresholding the low-SNR parameters is equivalent to assuming a delta-function prior over them. This collapses the parameter space down to a lower dimensional subspace (Figure 2e). Because the stimulus subspace prior assumes a flat prior over the high-SNR parameters there are no further constraints on the parameters within the lower dimensional subspace (Figure 2e).

The MAP objective for a linear model under Gaussian noise and the stimulus subspace prior is

$$g_{MAP}(\boldsymbol{\beta}') = (\mathbf{Y} - \mathbf{XQ}\boldsymbol{\beta}')^2 + \sum_{j,\delta_j > \varepsilon} 0 \cdot \beta'^2_j + \sum_{j,\delta_j \leq \varepsilon} \infty \cdot \beta'^2_j. \quad (0.27)$$

The minima of the MAP objective (0.27) can also be obtained via the LGG solution (0.11). And the MAP estimate of the parameters can be written as

$$\boldsymbol{\beta}^* = \left[\mathbf{Q}\mathbf{P}_\varepsilon \mathbf{Q}^T\right]^+ \mathbf{X}^T \mathbf{Y}, \quad (0.28)$$

where $\mathbf{P}_\varepsilon$ is as defined by equation (0.26), and the superscript-plus denotes the pseudo-inverse. Like the covariance subspace prior, the proper value of the noise threshold, $\varepsilon$, can be determined by cross validation. The stimulus subspace prior has been applied successfully in several neuronal system identification studies (David and Gallant, 2005; David et al., 2004; Theunissen et al., 2001; Woolley et al., 2006).

*3.7 Smooth priors and automatic smoothness determination*

None of the Gaussian priors considered so far are motivated biophysically. The flat prior, the spherical Gaussian prior and the independence prior were introduced for computational convenience. Even the correlated Gaussian priors, such as the stimulus covariance prior and the subspace priors, were developed merely to increase the accuracy of the parameter estimate by noise reduction. However, neurophysiological studies over the years have accumulated findings that can be incorporated into appropriate priors. For example, we might expect the stimulus-response mapping function to be a smooth function of space and time (Daugman, 1980; Simoncelli and Heeger, 1998). This knowledge can be incorporated into a neuronal system identification algorithm by assuming a *smooth prior*.

The smooth prior can also be formulated as a Gaussian prior. This is achieved by replacing $\mathbf{A}^{-1}$ in equation (0.8) by an appropriate differential operator that measures local smoothness. These operators include the Laplacian, $\nabla_x^2$ (Smyth et al., 2003; Willmore and Smyth, 2003), and the pair-wise distance operator that gives the square distance between all pairs of parameter values (Sahani and Linden, 2003). For example, using the Laplacian as a measure of local smoothness, the smooth prior can be written in the form of a Gaussian prior,

$$p(\boldsymbol{\beta}) \sim \exp\left[-\tfrac{1}{2\delta^2}\boldsymbol{\beta}^T \nabla_x^2 \boldsymbol{\beta}\right]. \tag{0.29}$$

Here $1/\delta$ is the length scale over which we expect the mapping function to be smooth, relative to the discretization of the stimulus. Under this smooth prior, the MAP objective function is

$$\begin{aligned} g_{MAP}(\boldsymbol{\beta}) &= (\mathbf{Y}-\mathbf{X}\boldsymbol{\beta})^2 + \tfrac{\sigma^2}{\delta^2}(\nabla_x\boldsymbol{\beta})^T(\nabla_x\boldsymbol{\beta}) \\ &= (\mathbf{Y}-\mathbf{X}\boldsymbol{\beta})^2 + \lambda \|\nabla_x\boldsymbol{\beta}\|_2^2, \end{aligned} \tag{0.30}$$

where $\lambda = \sigma^2/\delta^2$.

The smooth prior induces a form of regularization that penalizes the least-squares solution by the $\ell_2$-norm of the gradient of the model parameter. Rather than forcing the values of model parameters to be small as in ridge regression (§3.3), the smooth prior imposes smoothness by forcing the local changes in the parameter to be small. The amount of smoothness is governed by the choice of the differential operator. Once the operator is chosen, smoothness can still be controlled by the hyperparameter, $\lambda$. If $\lambda$ is large, the MAP objective (0.30) will be dominated by the regularizer. This will force the optimization algorithm to minimize the regularizer to give a very smooth mapping function. As the value of $\lambda$ decreases, the MAP objective will eventually be dominated by the loss functional, since the regularizer is multiplied by a tiny $\lambda$. This will force the optimization algorithm to minimize the loss functional to fit the data. In this case, the MAP estimate will produce a less smooth mapping function.

In practice, a finite difference matrix, $\mathbb{D}^2$, is used to approximate the $\nabla_x^2$ operator. Consequently, the MAP estimate under the smooth prior (0.29) can also be derived from the LGG solution (0.11)

$$\boldsymbol{\beta}^* = \left(\mathbf{X}^T\mathbf{X} + \lambda \mathbb{D}^2\right)^{-1}\mathbf{X}^T\mathbf{Y}.$$

Because smoothness is controlled by one hyperparameter, $\lambda$, the optimal smoothness can be determined by cross-validation. Alternatively, $\lambda$ can be determined by evidence maximization. This approach chooses the value of $\sigma^2$ and $\delta^2$ that maximizes the evidence as in ARD (§3.4.1). However, because it is used to determine the optimal level of smoothness, it is called automatic smoothness determination (Sahani and Linden, 2003). Alternatively, by introducing ordering to the features the *fused lasso* method can be applied as a smooth prior. In this case the differences between neighboring features are penalized by a second penalty term (Tibshirani et al., 2005).

*3.8 The spherical Laplace prior, the sparse prior and boosting*

All the priors that we have discussed so far are Gaussian priors. Although Gaussian priors offer many computational advantages, they are not the only useful priors. For example, to aid visualization and interpretation of the estimated mapping function, it is often desirable to obtain a sparse model that has few non-zero parameters. A *sparse prior* can be used to bias the model toward sparseness. A sparse prior is often defined by a prior distribution that is more sharply peaked at zero than is the MVG. The sharp peak at zero reflects the strong belief that all components of $\boldsymbol{\beta}$ should be zero. When the peak of the prior located at zero is infinitely sharp it becomes a delta-function, which pins all the parameter values to zero. This will give the ultimate sparse model, but the model will have no predictive power and so will be useless. A more useful sparse prior is the spherical Laplace prior,

$$p(\boldsymbol{\beta}) \propto \exp\left[\frac{-1}{b_{\boldsymbol{\beta}}}\|\boldsymbol{\beta}\|_1\right]. \qquad (0.31)$$

Here $b_{\boldsymbol{\beta}}$ is the scale parameter for a zero-mean Laplace distribution with variance $2b_{\boldsymbol{\beta}}^2$, and $\|\boldsymbol{\beta}\|_1 \equiv \sum_{j=1}^{d}|\beta_j|$ is the $\ell_1$-norm of the parameter vector, $\boldsymbol{\beta}$.

Although the spherical Laplace prior is not a Gaussian prior, as long as the noise distribution is assumed to be Gaussian, the MAP objective will still be a penalized least-squares problem. Under the spherical Laplace prior (0.31), the MAP objective of the linear model is

$$\begin{aligned} g_{MAP}(\boldsymbol{\beta}) &= (\mathbf{Y}-\mathbf{X}\boldsymbol{\beta})^2 - 2\sigma^2\log(p(\boldsymbol{\beta})) \\ &= (\mathbf{Y}-\mathbf{X}\boldsymbol{\beta})^2 - 2\sigma^2\left[\frac{-1}{b_{\boldsymbol{\beta}}}\sum_{j=1}^{d}|\beta_j|\right] \\ &= (\mathbf{Y}-\mathbf{X}\boldsymbol{\beta})^2 + \lambda\|\boldsymbol{\beta}\|_1 \end{aligned} \qquad (0.32)$$

Equation (0.32) resembles the ridge regression objective (0.14), but it is a different penalized least-squares problem called the *lasso* (Tibshirani, 1996). The lasso objective (0.32) is regularized by the $\ell_1$-norm rather than the $\ell_2$-norm of $\boldsymbol{\beta}$. Because the gradient of the $\ell_1$-norm regularizer does not exist everywhere, the general solution of the lasso cannot be solved via calculus as in the case of Gaussian priors. In general, the lasso has no closed-form solution. This makes computing the MAP estimate difficult. Efficient algorithms for minimizing the lasso have been developed over the last 15 years. These include the homotopy algorithm (Osborne, 2000), least angle regression [LARS, (Efron et al., 2004)], fast iterative shrinkage-thresholding algorithm [FISTA, (Beck and Teboulle, 2009)], and coordinate descent (Friedman et al., 2010)]. Inspection of these algorithms shows why the lasso solution is sparse. Because the $\ell_1$-norm regularizer is a sum of the absolute value of the model parameters, it is not differentiable whenever a parameter takes the value zero. The locations where the derivative of a function is not defined are called *corner points* (Figure 3). From the theory of optimization, whenever the objective function contains these corner points the solution will be sparse (Tibshirani, 1996).

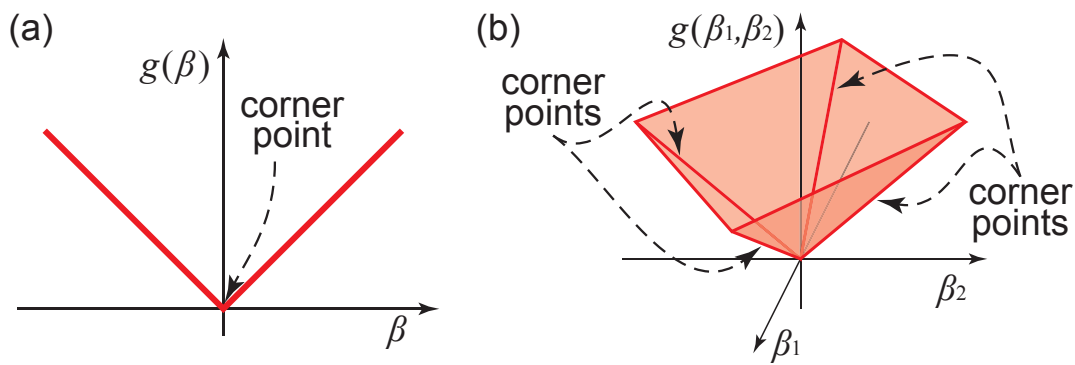

**Figure 3: Corner points.** Corner points are locations where the function's derivative does not exist. (a) The regularizer of a 1-dimensional Laplace prior has a corner point at $\beta = 0$. (b) The regularizer of 2-dimensional Laplace prior has corner points all along the axes of the parameter space. For higher-dimensional Laplace priors, the corner points are on the hyperplanes defined by $\beta_i = 0$.

3.8.1 ε-Boosting with early stopping and the lasso

$\varepsilon$-Boosting is an iterative algorithm for fitting the parameters, $\boldsymbol{\beta} = \left[\beta_1, \beta_2, \cdots, \beta_d\right]^T$, of any model that is linear in $\boldsymbol{\beta}$ (Buhlmann and Yu, 2003; Friedman, 2001). These include linear models (§3) and linearized models (§4.1). The essence of the $\varepsilon$-boosting algorithm is to build up the model iteratively by refitting the model residual from the previous $\varepsilon$-boosting iteration. The model residual is the difference between the measured response and the response predicted by the model. Because all $\beta_j$ are initialized to zero, the initial model residual is equal to the measured response. During each $\varepsilon$-boosting iteration, the single feature, $x_j$, is identified such that it is most correlated with the residual. Then the corresponding parameter, $\beta_j$, for $x_j$ is increased by a small step, $\varepsilon$. This slight change in $\boldsymbol{\beta}$ will result in a different model prediction, and will therefore change the model residual for the next round of $\varepsilon$-boosting.
Early stopping is a regularization technique that is commonly used during neural network training to prevent overfitting (Lau et al., 2002; Lehky et al., 1992; Sarle, 1995). However, this regularization technique can be applied to other iterative algorithms, such as gradient descent and $\varepsilon$-boosting. To apply early stopping in $\varepsilon$-boosting, the model residual at each iteration is monitored, and the $\varepsilon$-boosting algorithm is terminated when the residual reaches the noise level of the data. This stopping criterion is usually determined by cross-validation. In practice, the noise level is often reached before all the $\beta_j$ take their initial step. Therefore, many components of $\boldsymbol{\beta}$ will likely be zero when $\varepsilon$-boosting is stopped. As a result, the estimate of $\boldsymbol{\beta}$ is usually very sparse. Consequently $\varepsilon$-boosting with early stopping offers a simple way to obtain a sparse estimate of $\boldsymbol{\beta}$ (Zhang and Yu, 2005). This method has been successfully applied to neuronal system identification of visual neurons (Willmore et al., 2005) and auditory neurons (David et al., 2007).

To see how $\varepsilon$-boosting with early stopping is related to the lasso, consider the following. When $\varepsilon$-boosting is stopped early, the model parameters, $\beta_j$, are prevented from reaching their final values. Because $\beta_j$ will typically grow as $\varepsilon$-boosting iterates, early stopping effectively imposes a constraint on the norm of $\boldsymbol{\beta}$. Recent research in statistics has demonstrated that $\boldsymbol{\beta}$ obtained by early-stopped $\varepsilon$-boosting is close to the solution to the lasso for different values of the regularization constant, $\lambda$ (Rosset et al., 2004; Zhao and Yu, 2007). If $\varepsilon$-boosting is stopped during the early iterations where only a few $\beta_j$ have been fitted, $\|\boldsymbol{\beta}\|_1$ will still be relatively small. The resulting $\boldsymbol{\beta}$ will approximate the solution of the lasso for a large value of $\lambda$, which penalizes large $\|\boldsymbol{\beta}\|_1$ severely. As $\varepsilon$-boosting proceeds, $\|\boldsymbol{\beta}\|_1$ will increase. If $\varepsilon$-boosting is stopped at this point, the resulting $\boldsymbol{\beta}$ will closely resemble the lasso solution for a smaller value of $\lambda$, where $\|\boldsymbol{\beta}\|_1$ is penalized less severely. Hence, stopping the $\varepsilon$-boosting algorithm early affects the MAP estimate of $\boldsymbol{\beta}$ the same way that choosing the value of $\lambda$ affects the lasso. However, $\lambda$ is a nontrivial function of the training data, so there is no simple formula for $\lambda$ that corresponds to the early stopping of the $\varepsilon$-boosting algorithm.

## 4 Parametric nonlinear models

Up to this point we have considered only linear models. In practice, linear models are often too restrictive and they cannot describe the nonlinear behaviors of neurons accurately. Fortunately, the MAP framework is equally applicable to nonlinear model classes. Nonlinear models can be classified as parametric or nonparametric. The model complexity of *parametric models* is fixed and independent of the data, while the complexity of *nonparametric models* is data-dependent. In this section, we explore parametric nonlinear models; nonparametric nonlinear models will be discussed in the next section (§5).

The model class for parametric nonlinear models contains nonlinear mapping functions, $f_\theta$, that are indexed by the parameter vector, $\boldsymbol{\theta}$. Two nonlinear parametric model classes have been used in neuronal system identification: *linearized models* (§4.1) and *second-order models* (§4.2). Although the noise distribution is assumed Gaussian in both cases, the model classes are different. Because the likelihood depends on both the noise distribution and the model classes, the likelihood, and therefore the loss functional, will also differ. In this section we only survey the published regularization techniques for the respective models. But in theory, any of the priors (and regularization techniques) discussed in section §3 may be used with the nonlinear models in this section.

*4.1 Linearized models and linearized reverse correlation*

Linearized models are characterized by a front-end nonlinear transformation that attempts to describe the nonlinearity of the neuronal system. This transformation is called the linearizing transform, and it is denoted by $\mathbb{L}$. Because linearized models try to explicitly mimic the nonlinearities applied to the stimulus, these models often require substantial biophysical knowledge of the system. If the linearizing transform, $\mathbb{L}$, is able to describe the nonlinearities of the neuronal system then the relationship between the nonlinearly transformed stimulus, $\mathbb{L}(\mathbf{x})$, and the response, $y$, will be more linear than the relationship between $\mathbf{x}$ and $y$. In this case, the relationship between **x** and $y$ is said to be linearized by $\mathbb{L}$. Therefore, a linearized model is nothing more than a linear model of the nonlinear features.

The model class of linearized models consists of nonlinear mapping functions of the form

$$f_\theta(\mathbf{x}) = \mathbb{L}(\mathbf{x})^T \boldsymbol{\beta}. \tag{0.33}$$

Each model is indexed by $\theta = \boldsymbol{\beta}$, a vector of linear coefficients of the nonlinear features. Because the mapping function (0.33) is completely linear after the linearizing transform, the MAP objective resembles equation (0.10), except that **x** is replaced by $\mathbb{L}(\mathbf{x})$. This substitution can be achieved by defining the *stimulus feature matrix*,

$$\mathbf{L}_\mathbf{X} = \left[ \mathbb{L}(\mathbf{x}_1), \cdots, \mathbb{L}(\mathbf{x}_N) \right]^T. \tag{0.34}$$

The matrix $\mathbf{L}_\mathbf{X}$ can be viewed as the stimulus matrix in the feature space. Hence, the MAP objective for a linearized model under Gaussian noise and Gaussian prior with zero mean resembles equation (0.10), except that **X** is replaced with $\mathbf{L}_\mathbf{X}$,

$$g_{MAP}(\boldsymbol{\beta}) = (\mathbf{Y} - \mathbf{L}_\mathbf{X} \boldsymbol{\beta})^2 + \sigma^2 \boldsymbol{\beta}^T \mathbf{A}^{-1} \boldsymbol{\beta}. \tag{0.35}$$

Likewise, the MAP estimate, $\boldsymbol{\beta}^*$, takes the same form as the LGG solution (0.11) with **X** replaced by $\mathbf{L}_\mathbf{X}$,

$$\boldsymbol{\beta}^* = \left( \mathbf{L}_\mathbf{X}^T \mathbf{L}_\mathbf{X} + \sigma^2 \mathbf{A}^{-1} \right)^{-1} \mathbf{L}_\mathbf{X}^T \mathbf{Y}. \tag{0.36}$$

Algorithms that use linearized models and equation (0.36) for neuronal system identification are collectively referred to as *linearized reverse correlation*. These algorithms are often implemented in two stages. The first stage computes the stimulus feature matrix, $\mathbf{L}_\mathbf{X}$, in (0.34). Because $\mathbf{L}_\mathbf{X}$ depends only on the stimulus and can be computed before the response is acquired this is often implemented as a preprocessing stage of the algorithm. The second stage then computes the value of $\boldsymbol{\beta}^*$ via equation (0.36).

There is one important difference between the linearized and linear models that deserves special attention. The Gaussian prior $p(\boldsymbol{\beta})$ for the linearized model is defined over the model parameters in the feature space, not the stimulus space. Because $\mathbb{L}$ is a nonlinear transformation, a Gaussian prior in the feature space may not be a Gaussian in the stimulus space. Consequently, the region of high prior probability may no longer be a Gaussian ellipsoid.

### 4.1.1 The linearizing transforms

In order to carry out the preprocessing stage of linearized reverse correlation, it is necessary to know the exact form of the linearizing transform, $\mathbb{L}$. A suitable choice of $\mathbb{L}$ should be one that closely mimics the nonlinear functional properties of the system. However, the nonlinear properties of many neuronal systems are unknown. Therefore, linearized reverse correlation is often best applied to well-studied systems where prior knowledge can be exploited to aid the construction of an appropriate $\mathbb{L}$. Here we describe several biologically inspired linearizing transforms that have been applied to neuronal SI.

The first study that applied a linearized model aimed to explain the spectral power nonlinearity in the auditory neuron (Aertsen and Johannesma, 1981a). To describe non-phase-locking neurons in the auditory midbrain, the spectrogram linearizing transform was developed. The spectrogram is the temporally localized Fourier power of the stimulus. Therefore it is also called the dynamic power spectrum (Aertsen and Johannesma, 1980), and can be written as

$$\mathbb{L}(\mathbf{x}_i) \equiv \left| \mathbb{F}_1 \left[ x(t) \cdot W_1(t - i\Delta t) \right] \right|^2. \tag{0.37}$$

Here $x(t)$ is the sound stimulus represented by the pressure wave. Temporal localization is achieved by a one-dimensional window function, $W_1(t - i\Delta t)$ (usually a Gaussian or a Hanning window), centered at the $i$th sample, where $\Delta t$ is the sampling interval, and the notation $\mathbb{F}_1[\cdot]$ represents the one-dimensional discrete Fourier transform. The spectrogram accounts for a ubiquitous nonlinearity in the early stage of auditory processing. Consequently, the spectrogram has become the default linearizing transform for many auditory system identification algorithms (Aertsen and Johannesma, 1981a; Kowalski et al., 1996; Machens et al., 2004; Theunissen et al., 2000).

Linearized reverse correlation was also used to describe the phase invariant nonlinearity of V1 complex cells (David et al., 2004; Nishimoto et al., 2005; Nykamp and Ringach, 2002). Two different linearizing transforms have been developed to account for the phase invariant nonlinearity of V1 complex cells (Movshon et al., 1978; De Valois et al., 1982). The first linearizing transform is the Fourier power transform that transforms the stimulus into the frequency domain while discarding the phase information (David and Gallant, 2005). This linearizing transform can be written as

$$\mathbb{L}(\mathbf{x}) \equiv \left| \mathbb{F}_2(\mathbf{x}) \right|^2, \tag{0.38}$$

where $\mathbf{x}$ represents an image and $\mathbb{F}_2[\cdot]$ is the two-dimensional discrete Fourier transform. However, V1 also contains phase sensitive simple cells. For these neurons, the Fourier power transform (0.38) is not appropriate. Consequently, the phase-separated Fourier transform was derived to describe both simple and complex cells in V1 (David et al., 2004). This linearizing transform is able to account for phase invariance while maintaining all the phase information. The phase-separated Fourier transform can be written as

$$\mathbb{L}(\mathbf{x}) = \left[ \left|\text{Re}\left[\mathbb{F}_2(\mathbf{x})\right]\right|_+, \left|\text{Im}\left[\mathbb{F}_2(\mathbf{x})\right]\right|_+, \left|\text{Re}\left[\mathbb{F}_2(\mathbf{x})\right]\right|_-, \left|\text{Im}\left[\mathbb{F}_2(\mathbf{x})\right]\right|_- \right]^T, \quad (0.39)$$

where $\text{Re}[\cdot]$ and $\text{Im}[\cdot]$ denote the real and imaginary parts, and $|\cdot|_+$ and $|\cdot|_-$ denote positive and negative rectification.

One limitation of the Fourier power and the phase-separated Fourier transform is that they are global with respect to the stimulus domain of the mapping function (e.g. the classical receptive field). Therefore, they will average out the effect of any local stimulus structure (e.g. local spatial features within the receptive field) on the neuron's response. To overcome this limitation, the second linearizing transform, the local spectral reverse correlation (LSRC) was developed (Nishimoto et al., 2006). The LSRC describes the spatially localized structures of the mapping function. LSRC is also a linearized model where the linearizing transform is the spatially localized Fourier power of the stimulus. The corresponding linearizing transform can be written as

$$\mathbb{L}(\mathbf{x}) \equiv \left| \mathbb{F}_2\left[ \mathbf{x} \circ \mathbf{W}_2(x_1 - j\Delta x, x_2 - k\Delta x) \right] \right|^2. \quad (0.40)$$

Here $\mathbf{W}_2(x_1 - j\Delta x, x_2 - k\Delta x)$ is an overlapping tiling of a localized two-dimensional Gaussian window and $\Delta x$ is the spacing between each tiling. The circle multiplication symbol, $\circ$, denotes the matrix Hadamard product (i.e. element-by-element or entry-wise product). The linearizing transform in equation (0.40) is conceptually similar to the spectrogram (0.37) used in auditory SI, except that features in LSRC are the localized Fourier channels in space rather than time.

As mentioned earlier, the linearizing transform may be applied to the stimulus before acquiring the response. Hence, some researchers do not treat these preprocessing steps as part of the system identification algorithm. However, many preprocessing procedures that are applied to the stimulus can be viewed as linearizing transforms. For example, the stimulus is often normalized to have the same mean luminance and contrast to mimic the normalization that occurs in early sensory processing (David et al., 2004). This procedure can be viewed as a linearizing transform of the stimulus, hence are part of the system identification algorithm.

Another class of commonly used linearizing preprocessing steps is to map the stimulus into a space where the relevant stimulus features are defined explicitly. These include PCA, independent component analysis (ICA), and wavelet decomposition of the stimulus (Prenger et al., 2004; Willmore et al., 2008). Although these transformations are linear, a static nonlinearity is typically applied to the linear features. Commonly used static nonlinearities include rectification, saturation (compressive nonlinearity), squaring (expansive nonlinearity), and thresholding (Willmore et al., 2008). In addition, data driven linearizing transforms have been successfully applied to neural data in recent years (Park and Pillow, 2011; Park et al., 2013; Rapela et al., 2010). In general, a series of preprocessing steps can be build together to create complex linearizing transforms.

### 4.1.2 Regularization of linearized models

A linearized model is a linear model in a nonlinear feature space. Therefore, all priors and methods of regularization used with linear models can be applied directly to the linearized models. In auditory system identification, early spectrogram-linearized models were not regularized, so they implicitly assumed a flat prior (Aertsen and Johannesma, 1981a; Eggermont et al., 1983b). Subsequently, many of the priors and regularization heuristics presented in section §3 came into use with these models. These include the spherical Gaussian prior (Machens et al., 2004), the independence prior (Sahani and Linden, 2003), the smooth prior (Sahani and Linden, 2003), and the stimulus subspace prior (Theunissen et al., 2001; Woolley et al., 2006). In visual system identification, the Fourier power model and phase separated Fourier model have implicitly assumed a stimulus subspace prior (David and Gallant, 2005; David et al., 2004). LSRC was developed recently (Nishimoto et al., 2006), however, no regularization methods or priors have been tested yet.

4.1.3 Using natural stimulus with linearized models

In order to understand the function of sensory neurons under natural conditions it is necessary to use stimuli that mimic the neuron's natural stimulation (Aertsen and Johannesma, 1981b; David et al., 2004; Ringach et al., 2002; Theunissen et al., 2000; Touryan et al., 2005; Willmore and Smyth, 2003). Using natural stimuli with linearized models requires careful regularization to produce unbiased estimates of the mapping function. Because natural stimuli are low frequency biased, both the nonlinear features and the stimulus feature matrix, $\mathbf{L_X}$, will inherit this spectral bias. Hence, the feature autocovariance matrix, $\mathbf{L_X^T L_X}$ will have a wide range of eigenvalues $\delta_j$. (If the stimulus is white, or spectrally unbiased, all $\delta_j$ would have similar magnitudes.) When $\mathbf{L_X^T L_X}$ is diagonalized as $\mathbf{Q_L D Q_L^T}$ with $\mathbf{D} = diag(\delta_j)$, the low-frequency features will have large $\delta_j$, and the high-frequency features will have small $\delta_j$. Consequently, the model parameters of the high-frequency features will have low SNR and will not be estimated accurately without proper regularization.

To properly regularize a linearized model that is estimated with natural stimuli it is advantageous to use a prior that incorporates some information about the statistics of the stimulus. These *stimulus-dependent priors* include the stimulus covariance prior (§3.5, Figure 2c), the stimulus subspace prior (§3.6.2, Figure 2e), and the covariance subspace prior (§3.6.1, Figure 2d). For example, if the stimulus covariance prior is assumed, its Wiener filtering property (0.22) will effectively suppress noise in the high-frequency features. Using the LGG solution (0.11) and equation (0.22), we can derive the MAP estimate for a linearized model under the stimulus covariance prior:

$$\begin{aligned} \boldsymbol{\beta}^* &= \left[ \mathbf{L}_\mathbf{X}^T \mathbf{L}_\mathbf{X} + \sigma^2 \left( \mathbf{L}_\mathbf{X}^T \mathbf{L}_\mathbf{X} \right)^{-1} \right]^{-1} \mathbf{L}_\mathbf{X}^T \mathbf{Y} \\ &= \mathbf{Q}_\mathbf{L} \left[ diag(w_j) \right] \mathbf{Q}_\mathbf{L}^T \mathbf{L}_\mathbf{X}^{-1} \mathbf{Y} . \end{aligned} \quad (0.41)$$

Here $w_j = \delta_j^2 / (\delta_j^2 + \sigma^2)$, and the eigenvalues, $\delta_j$, represent the relative spectral power of the features in the stimulus. Natural stimuli tend to have a 1/f amplitude spectrum, and the high-spectral-power features ($\delta_j \gg \sigma^2$) that have strong signals in the data are the low-frequency features. Because $w_j \approx 1$ when $\delta_j \gg \sigma^2$, model parameters for the low-frequency features will only shrink slightly. In contrast, the low-spectral-power features ($\delta_j \ll \sigma^2$) of natural stimuli are the high-frequency features. Because $w_j \ll 1$ when $\delta_j \ll \sigma^2$, model parameters for the high-frequency features will be attenuated significantly. These low-SNR parameters can be removed completely if desired by assuming either the stimulus subspace prior or the covariance subspace prior (§3.6).

Note that the stimulus-dependent priors will only attenuate noise when the stimulus is spectrally biased as occurs with natural stimuli. If the stimuli have a flat power spectrum (i.e. they are white), then the feature autocovariance matrix, $\mathbf{L}_\mathbf{X}^T \mathbf{L}_\mathbf{X}$, in equation (0.41) will be proportional to the identity matrix. In this case, equation (0.41) will reduce to the ridge regression solution (0.15), where the regularizer is simply the $\ell_2$-norm of the parameter vector.

*4.2 Second-order Wiener-Volterra models and spike-triggered covariance*

Linearized models are not useful when knowledge is insufficient to form an appropriate linearizing transform. In this case, the nonlinear Wiener-Volterra models provide a general alternative to linearized models. In theory, any finite-memory nonlinear system can be expanded as a Volterra series (Boyd and Chua, 1985; Volterra, 1959) or its orthogonalized form, the Wiener series (Eggermont, 1993; Wiener, 1958). The linear model can then be viewed as the first-order term of this infinite series, and nonlinear Wiener-Volterra models can be constructed by including higher-order terms. Because the number of terms in the model can increase indefinitely as more data become available, model complexity can grow to fit the data. Therefore, the Wiener-Volterra series is a nonparametric model. However, because the data generated in experiments are usually limited most applications of Wiener-Volterra models to neuronal system identification are truncated to second order (Citron and Emerson, 1983; Eggermont, 1993; Emerson et al., 1987; Gaska et al., 1994; Lewis and van Dijk, 2004; Mancini et al., 1990; Victor and Shapley, 1980). This limits the complexity of the model and makes the model parametric. For this reason for the remainder of this review, we will treat Wiener-Volterra models as parametric and refer to them as second-order models.

The model class for second-order models consists of functions of the form

$$f_\theta(\mathbf{x}) = \mathbf{x}^T\boldsymbol{\beta} + \mathbf{x}^T\mathbf{B}\mathbf{x}, \tag{0.42}$$

where $\boldsymbol{\theta} = \{\boldsymbol{\beta}, \mathbf{B}\}$ are the model parameters. The vector $\boldsymbol{\beta}$ contains the linear coefficients, and the symmetric matrix $\mathbf{B}$ contains the quadratic coefficients. Under Gaussian noise and the flat prior, the MAP estimate for these model parameters can be obtained by minimizing the following MAP objective,

$$g_{MAP}(\boldsymbol{\beta}, \mathbf{B}) = \sum_{i=1}^{N} \left\{ y_i - \left[\mathbf{x}^T\boldsymbol{\beta} + \mathbf{x}^T\mathbf{B}\mathbf{x}\right] \right\}^2. \tag{0.43}$$

Non-flat priors may be used with the second-order model, but the regularizers induced by these priors are much easier to write down and understand when the model is re-parameterized. In the next section (§4.2.1) we will re-parameterize appropriately and present the full MAP objective and MAP estimate for the second-order model.

Because the second-order model does not require explicit knowledge of the nonlinearities in the system, this model class has been widely applied in system identification studies of early sensory systems. In the auditory system, second-order models have explained the envelope response in the papilla of bullfrogs (Yamada and Lewis, 1999), inferior colliculus of owls (Keller and Takahashi, 2000), A1 of ferrets (Kowalski et al., 1996), and auditory forebrain of songbirds (Sen et al., 2001). The second-order model has been used to describe frequency-specific modulations in the auditory nerve fiber of cats (Young and Calhoun, 2005). In the visual system, second-order models have explained phase invariance (Gaska et al., 1994; Touryan et al., 2002), directional selectivity (Citron and Emerson, 1983) and velocity selectivity (Emerson et al., 1987) of cat V1 complex cells.

### 4.2.1 Second-order model as a special form of linearized model

Although the second-order model (0.42) is nonlinear in the stimulus, $\mathbf{x}$, this model is linear in the model parameters, $\mathbf{\theta} = \{\mathbf{\beta}, \mathbf{B}\}$. Therefore computing the MAP estimate for $\mathbf{\beta}$ and $\mathbf{B}$ is still a linear estimation problem. For such an estimation problem, it is more convenient to rewrite equation (0.42) as

$$f_\theta(\mathbf{x}) = \mathbb{S}(\mathbf{x})^T \mathbf{\theta}. \tag{0.44}$$

Here $\mathbb{S}(\mathbf{x})$ is a nonlinear function of $\mathbf{x}$ that incorporates all the nonlinearities of equation (0.42), and $\mathbf{\theta}$ is a column vector of parameters consisting of the elements of $\mathbf{\beta}$ and $\mathbf{B}$. Because equation (0.42) contains both linear and quadratic terms in $\mathbf{x}$, the function $\mathbb{S}(\mathbf{x})$ must transform the stimulus into a set of first- and second-order features. Therefore

$$\mathbb{S}(\mathbf{x}) = [\underbrace{x_1, \cdots, x_d}_{\text{1st order features}} \mid \underbrace{x_1 x_1, \cdots, x_1 x_d \mid x_2 x_1, \cdots x_2 x_d \mid \cdots \mid x_d x_1, \cdots x_d x_d}_{\text{2nd order features}}]^T. \tag{0.45}$$

Here $d$ is the dimensionality of the stimulus, and the subscripts denotes components of the vector $\mathbf{x}$ rather than the sample index. The nonlinear transform $\mathbb{S}(\mathbf{x})$, as defined by equation (0.45) will subsume all the nonlinearities in the second-order model (0.42).

Comparing equation (0.44) and equation (0.33), it is clear that the second-order model is merely a special kind of linearized model, where the linearizing transform is $\mathbb{L}(\mathbf{x}) = \mathbb{S}(\mathbf{x})$. For this reason, $\mathbb{S}(\mathbf{x})$ is also called the second-order linearizing transform. To use the same formulae that is derived for computing the MAP estimate of linearized models, we need to define the stimulus feature matrix,

$$\mathbf{S_X} = \left[ \mathbb{S}(\mathbf{x}_1), \cdots, \mathbb{S}(\mathbf{x}_N) \right]^T.$$

Now, if the noise and the prior are both assumed to be Gaussian, the MAP objective will look like equation (0.35) with $\mathbf{L_X}$ and $\boldsymbol{\beta}$ replaced by $\mathbf{S_X}$ and $\boldsymbol{\theta}$ respectively,

$$g_{MAP}(\boldsymbol{\theta}) = (\mathbf{Y} - \mathbf{S_X}\boldsymbol{\theta})^2 + \sigma^2 (\boldsymbol{\theta} - \boldsymbol{\theta}_0)^T \mathbf{A}^{-1} (\boldsymbol{\theta} - \boldsymbol{\theta}_0).$$

Likewise, the MAP estimate may be obtain via the LGG solution (0.11), and it will resemble equation (0.36),

$$\boldsymbol{\theta}^* = \left( \mathbf{S_X}^T \mathbf{S_X} + \sigma^2 \mathbf{A}^{-1} \right)^{-1} \mathbf{S_X}^T \mathbf{Y}. \qquad (0.46)$$

Subsequently, the parameters in $\boldsymbol{\theta}^*$ may be rearranged and divided into the MAP estimate for the linear ($\boldsymbol{\beta}^*$) and quadratic ($\mathbf{B}^*$) coefficients.

### 4.2.2 Spike-triggered covariance

Spike-triggered covariance (STC) is an algorithm for computing and interpreting the second-order model of a neuron's mapping function. In contrast to STA, STC computes the covariance matrix rather than the average of the spike-triggered stimulus ensemble. The *significant eigenvectors* of the spike-triggered covariance matrix will then characterize the relevant stimulus subspace of the neuronal system. For nonlinear neurons, several significant eigenvectors are typically recovered.

STC was originally used to approximate the response-conditioned distribution of the stimulus with a MVG, and then using this approximation to estimate its information content (de Ruyter van Steveninck and Bialek, 1988). Because the MVG is fully characterized by its first- and second-order statistics, the underlying model class for STC is a second-order model. However, because STC is simple to implement and easy to visualize and interpret, it has become a popular tool for neuronal system identification (Brenner et al., 2000; Felsen et al., 2005; Rust et al., 2005; Schwartz et al., 2002; Touryan et al., 2002).

In practice, STC is usually used in conjunction with white noise stimuli and the flat prior. When white noise is used, the feature autocovariance matrix, $\mathbf{S}_\mathbf{X}^T \mathbf{S}_\mathbf{X}$, is proportional to the identity. And under the flat prior assumption, $\mathbf{A}^{-1} = \mathbf{0}$. Therefore, equation (0.46) for the MAP estimate of $\boldsymbol{\theta}$ reduces to the ordinary spike-triggered solution,

$$\boldsymbol{\theta}^* = \mathbf{S}_\mathbf{X}^T \mathbf{Y}. \tag{0.47}$$

Equation (0.47) indicates the MAP estimate is simply the response weighted average of the features in the stimulus feature matrix, $\mathbf{S}_\mathbf{X}$. Because $\mathbf{S}_\mathbf{X}$ contains nonlinear second-order features in addition to the linear first-order features, the parameter vector, $\boldsymbol{\theta}^*$, contains both linear ($\boldsymbol{\beta}^*$) and quadratic ($\mathbf{B}^*$) coefficients. The response weighted average of the first-order features is simply the STA, and they are stored in the components of $\boldsymbol{\theta}^*$ that correspond to $\boldsymbol{\beta}^*$. Likewise, the response weighted average of the second-order features is precisely the STC, and they are stored in the components of $\boldsymbol{\theta}^*$ that correspond to $\mathbf{B}^*$.

To select the significant eigenvectors of STC, the eigenvalue of $\mathbf{B}^*$ is compared to the covariance fluctuations arising from random samples of the stimulus space. Eigenvectors whose corresponding eigenvalues are larger than these random fluctuations are considered significant. The relevant subspace of the neuronal system is spanned by the set of significant eigenvectors of $\mathbf{B}^*$. Because $\mathbf{B}^*$ is also the second-order kernel of a Wiener-Volterra model (0.42), STC merely characterizes this relevant subspace by the significant eigenvectors of the second-order kernel.

### 4.2.3 Computational issues in second-order models

Although the second-order model can be viewed as a linearized model, there are some important differences. First, the second-order model suffers more severely from the *curse of dimensionality* than the linearized models do. This means the number of model parameters for the second-order model will grow more rapidly as a function of the stimulus dimensionality when compared to the linearized model. This will in turn increase the data requirements for the second-order models. Recall that the second-order linearizing transform, $\mathbb{S}(\mathbf{x})$ in equation (0.45) , maps a $d$-dimensional stimulus, to a vector of $\sim (d^2 + d)$ features. Because there is one model parameter for each stimulus feature, the number of parameters is a quadratic function of the stimulus dimensionality. Compared to the second-order model, the growth in the number of parameters is milder for linearized models. For example, the number of parameters is $\frac{1}{2}d$ for the Fourier power transform and $2d$ for the phase-separated Fourier transform. In both case, the number of parameters is only a linear function of the stimulus dimensionality.

Second, the curse of dimensionality also limits the type of stimuli that can be used with second-order models. Because the size of the feature autocovariance matrix, $\mathbf{S}_\mathbf{X}^T \mathbf{S}_\mathbf{X}$, is $(d^2 + d) \times (d^2 + d)$ for a $d$-dimensional stimulus, $\mathbf{S}_\mathbf{X}^T \mathbf{S}_\mathbf{X}$ in equation (0.46) can become intractably large when $d$ is large. For example, a typical naturalistic stimulus may have $d = 1000$. The number of elements in $\mathbf{S}_\mathbf{X}^T \mathbf{S}_\mathbf{X}$ will be on the order of $d^4 \sim 10^{12}$. Inverting such a large matrix is well beyond the capacity of common computing platforms. This limitation precludes the application of (0.46) to correlated stimuli, which require normalization by the inverse of $\mathbf{S}_\mathbf{X}^T \mathbf{S}_\mathbf{X}$ to obtain an unbiased estimate of the model (§3.2).

Several experimental manipulations and computational methods have been developed to alleviate the curse of dimensionality for the second-order model. The simplest of these manipulations is to use low dimensional stimuli (Rust et al., 2005; Touryan et al., 2002). When the stimulus dimensionality, $d$, is small, the number of model parameters, $(d^2 + d)$, may be within the reach of the sample size, $N$. Because at least one data sample is required to accurately estimate the value of each model parameter, it is crucial to have $N \geq (d^2 + d)$. Furthermore, if $d$ is low enough, it is possible to use correlated stimuli with the second-order model. For a very small value of $d$, even $d^4$ may still be less than $N$, therefore the inverse of $\mathbf{S}_\mathbf{X}^T \mathbf{S}_\mathbf{X}$ may be computed accurately without numerical instability to correct for the bias introduced by stimulus correlation. The second manipulation involves applying a dimensionality reduction algorithm (such as principal component analysis) to the complex naturalistic stimulus in order to lower $d$ (Prenger et al., 2004). The third manipulation that enables the use of natural stimulus with the second-order model is to pre-whiten the stimulus. This involves normalizing each Fourier components to produce a flat spectrum (Simoncelli and Olshausen, 2001), or convolving a center-surround filter in the pixel domain to accentuate the high frequency features (e.g. edges). In either case, these calculations operate on objects with dimensionality $d$, which are computationally very feasible. This will ensure $\left(\mathbf{S}_\mathbf{X}^T \mathbf{S}_\mathbf{X}\right)^{-1} \propto \mathbf{I}$, and circumvent the direct inversion of $\mathbf{S}_\mathbf{X}^T \mathbf{S}_\mathbf{X}$. Although the higher-order correlations within natural stimuli may still introduce some estimation bias, in practice, they usually do not significantly alter the estimated parameters (Touryan et al., 2005). There are also computational algorithms that can circumvent the use of equation (0.46) and therefore by-pass the inversion of $\mathbf{S}_\mathbf{X}^T \mathbf{S}_\mathbf{X}$ completely. Recall that equation (0.46) is the closed-form solution for the extremum of the MAP objective (0.43). Since this MAP objective is a smooth and convex function, its extremum can be obtained efficiently by gradient-based optimization algorithms (Fletcher, 1987; Prenger et al., 2004). Alternatively, the MAP estimate can be computed via the method of recursive least-squares (Lesica et al., 2003; Stanley, 2002). Both approaches are more general than the application of equation (0.46), and they are applicable when closed-form solution of the MAP estimate is numerically difficult to compute.

## 5 Nonparametric nonlinear models

The previous section addressed parametric nonlinear models whose model complexity is fixed. However, parametric nonlinear models are limited in many ways. For example, the specific nonlinearities built into the linearized model can describe a highly nonlinear system, but constructing the linearizing transform requires substantial knowledge about the system that is often unavailable. A general nonlinearity in the second-order model can describe a mildly nonlinear system, but it requires much more data than the linearized models do. To model highly nonlinear systems with unknown nonlinearities, nonparametric nonlinear models are the natural choice. Because all nonparametric models in this review are nonlinear, we will simply refer to them as nonparametric models.

The MAP framework can also be extended to nonparametric model classes. In general, nonparametric model classes can be written as $\mathcal{M} = \{f_{\boldsymbol{\theta},\mathcal{D}}(\mathbf{x})\}$, where $f_{\boldsymbol{\theta},\mathcal{D}}$ is an arbitrary nonlinear function parameterized by a vector $\boldsymbol{\theta}$. The two subscripts $\mathcal{D}$ and $\boldsymbol{\theta}$ indicate that the data and the parameters are both needed to specify $f_{\boldsymbol{\theta},\mathcal{D}}$. The following section will focus on the unique aspects of a nonparametric nonlinear model class that assumes Gaussian noise: *kernel ridge regression*.

*5.1 Kernel regression model under Gaussian noise and Gaussian prior*

Kernel regression algorithms are the nonparametric analog of linearized reverse correlation and are universal function approximators (Hammer and Gersmann, 2003). However, kernel regression makes the nonlinear regression problem convex by converting it into a linear regression problem in a nonlinear feature space. Therefore, kernel regression is guaranteed to have a unique and globally optimal solution.

Since the inception of the first kernel regression algorithm in 1990s, many more have been developed in the machine learning and statistics community. This review will analyze two kernel regression algorithms: *support vector regression* (SVR) and *kernel ridge regression* (kRR). This section will focus on kernel regression algorithms that assume Gaussian noise and a Gaussian prior: kRR. The discussion of SVR is deferred to section §6.2, because SVR does not assume Gaussian noise.

Like linearized reverse correlation, kernel regression algorithms will first transform the stimulus, $\mathbf{x}$, into a high dimensional feature space by a feature map, $\Phi$. Then a linear model is fit to the set of nonlinear features, $\Phi(\mathbf{x})$. The model classes underlying kernel regression algorithms will be referred to as kernel regression models. These model classes consist of functions of the form

$$f_{\boldsymbol{\theta},\mathcal{D}}(\mathbf{x}) = \langle \Phi(\mathbf{x}), \boldsymbol{\beta} \rangle, \tag{0.48}$$

where $\langle \cdot, \cdot \rangle$ denotes an inner product in the feature space. From equation (0.48), it is clear that any kernel regression model is parameterized by the vector $\boldsymbol{\theta} = \boldsymbol{\beta}$. Although kernel regression models also depend on $\Phi$, $\Phi$ is a fixed nonlinear transform that is not fitted by the algorithm.

Because the feature space is usually a very high (possibly infinite) dimensional space, $\boldsymbol{\beta}$ may be an infinitely long vector. This prohibits the explicit computation of $\boldsymbol{\beta}$, which will in turn prohibits evaluating the model with equation (0.48). This problem is referred to as the *model representation problem*, and it is common to all kernel regression models. This problem will be addressed in section §5.1.2. For the time being, we will try to get the big picture without getting bogged down by the details. So we will temporarily ignore the model representation problem and proceed with the theory assuming a finite dimensional feature spaces. Under this assumption, the inner product, $\langle \cdot, \cdot \rangle$, in equation (0.48) may be replaced by the usual dot product, $\Phi(\mathbf{x})^T \boldsymbol{\beta}$ without loss of generality. Now, the kernel regression model will have the form of a linearized model (0.33), except the linearizing transform, $\mathbb{L}$, is replaced by the high dimensional feature map, $\Phi$.

Assuming both the noise and the prior are Gaussian, the MAP objective for the kernel regression model resembles equation (0.35),

$$g_{MAP}(\boldsymbol{\beta}) = (\mathbf{Y} - \boldsymbol{\Phi}_{\mathbf{X}} \boldsymbol{\beta})^2 + \sigma^2 (\boldsymbol{\beta} - \boldsymbol{\beta}_0)^T \mathbf{A}^{-1} (\boldsymbol{\beta} - \boldsymbol{\beta}_0). \tag{0.49}$$

Here $\boldsymbol{\Phi}_{\mathbf{X}} = \left[ \Phi(\mathbf{x}_1), \cdots, \Phi(\mathbf{x}_N) \right]^T$ is again the stimulus feature matrix. The MAP estimate for this class of models is easily obtained via the LGG solution (0.11),

$$\boldsymbol{\beta}^* = \left( \boldsymbol{\Phi}_{\mathbf{X}}^T \boldsymbol{\Phi}_{\mathbf{X}} + \sigma^2 \mathbf{A}^{-1} \right)^{-1} \boldsymbol{\Phi}_{\mathbf{X}}^T \mathbf{Y}. \tag{0.50}$$

Notice the striking similarity between equations (0.50), (0.36) and (0.46). The only difference between these equations is the stimulus feature matrices, $\boldsymbol{\Phi}_{\mathbf{X}}$, $\mathbf{L}_{\mathbf{X}}$, and $\mathbf{S}_{\mathbf{X}}$, respectively. This is because the kernel regression model, the linearized model, and the second-order model are all based on a linear model of nonlinearly transformed stimuli.

5.1.1 Regularization of kernel ridge regression

Because the parameter vector, $\boldsymbol{\beta}$, in kernel regression models contains a very large number of model parameters, kernel regression will overfit the data without a suitable prior. The simplest kernel regression algorithm is kernel ridge regression, and it assumes a spherical Gaussian prior just as in ordinary ridge regression (§3.3). Therefore the kRR MAP objective is similar to the ordinary ridge regression objective (0.14), but $\mathbf{X}$ is replaced by $\boldsymbol{\Phi}_\mathbf{X}$,

$$g_{MAP}(\boldsymbol{\beta}) = (\mathbf{Y} - \boldsymbol{\Phi}_\mathbf{X}\boldsymbol{\beta})^2 + \lambda \|\boldsymbol{\beta}\|_2^2. \tag{0.51}$$

Here $\lambda \equiv \sigma^2/\sigma_\boldsymbol{\beta}^2$ is again the regularization hyperparameter. The MAP estimate for $\boldsymbol{\beta}$ is readily obtained by setting $\mathbf{A} = \sigma_\boldsymbol{\beta}^2 \mathbf{I}$ in equation (0.50),

$$\boldsymbol{\beta}^* = \left( \boldsymbol{\Phi}_\mathbf{X}^T \boldsymbol{\Phi}_\mathbf{X} + \sigma^2 \underbrace{\sigma_\boldsymbol{\beta}^{-2}\mathbf{I}}_{\mathbf{A}^{-1}} \right)^{-1} \boldsymbol{\Phi}_\mathbf{X}^T \mathbf{Y}$$

$$= \left( \boldsymbol{\Phi}_\mathbf{X}^T \boldsymbol{\Phi}_\mathbf{X} + \lambda \mathbf{I} \right)^{-1} \boldsymbol{\Phi}_\mathbf{X}^T \mathbf{Y}. \tag{0.52}$$

Equation (0.52) is similar to the MAP estimate for ordinary ridge regression (0.15), except that $\mathbf{X}$ is replaced by $\boldsymbol{\Phi}_\mathbf{X}$.

As we showed for linearized models, the spherical Gaussian prior for kernel ridge regression is defined over the model parameters in the feature space, not the stimulus space. Because the feature map, $\boldsymbol{\Phi}$, is nonlinear, the region of high prior probability in the stimulus space may not be spherical or Gaussian. Therefore care must be taken when interpreting the prior for kernel ridge regression models.

One subtlety that distinguishes kRR from linearized models is the dimensionality of the feature space, $d_\mathcal{F}$. The dimensionality of the feature space for most linearized models is a constant multiple of the stimulus dimensionality, $d$, but $d_\mathcal{F}$ is usually much greater than $d$. Even though we can write the MAP estimate for the kRR parameters in equation (0.52), actually evaluating this expression could be a numerically intractable problem. This is because $d_\mathcal{F}$ may well surpass several billions, so the stimulus feature matrix, $\mathbf{\Phi_X}$, is a very large matrix of size $N \times d_\mathcal{F}$. The feature autocovariance matrix, $\mathbf{\Phi_X}^T \mathbf{\Phi_X}$, and $\lambda \mathbf{I}$ are gigantic matrices of size $d_\mathcal{F} \times d_\mathcal{F}$. Inversion of such huge matrices is practically impossible on most computing platforms. Moreover, $d_\mathcal{F}$ can be infinite! This brings us back to model representation problem that we have ignored until now.

5.1.2 Model representation and evaluation

Unlike any other model classes we have discussed so far, kRR models are not trivially represented. This means that even though a formula for computing the model parameters, $\boldsymbol{\beta}$, is derived in equation (0.52), in practice, we cannot actually compute or store $\boldsymbol{\beta}$. Without $\boldsymbol{\beta}$, the model representation in equation (0.48) is useless, and there is no way to evaluate the model response given a set of stimuli. This model representation problem is common to all kernel regression models. The solution to this problem lies in the so-called *kernel trick* (Aizerman et al., 1964; Boser et al., 1992; Schölkopf et al., 1998). Conceptually, the kernel trick converts the computation of an inner product in a feature space into an evaluation of a *kernel function*, $\kappa$, in the stimulus space.

For technical reason, we will not describe the theory behind the kernel trick, as it requires knowledge from functional analysis. Using the kernel trick, however, we can derive a numerically computable representation of the kernel ridge regression model (Appendix B):

$$f_{\theta,\mathcal{D}}(\mathbf{x}) = \sum_{i=1}^{N} \alpha_i \kappa(\mathbf{x}, \mathbf{x}_i). \tag{0.53}$$

Here $\alpha_i$ are the components of $\boldsymbol{\alpha} \equiv (\mathbf{K} + \lambda \mathbf{I})^{-1} \mathbf{Y}$, where $\mathbf{K}$ is the kernel matrix defined by the kernel function via $\mathbf{K}_{ij} \equiv \kappa(\mathbf{x}_i, \mathbf{x}_j)$. Equation (0.53) is completely defined by the kernel function, $\kappa$, and the data. Even $\alpha_i$ are defined by $\kappa$ through the kernel matrix, $\mathbf{K}$. Because $\kappa$ is computable for an infinite-dimensional feature space, the model represented by equation (0.53) will be computable irrespective of the feature space dimensionality.

Recall that nonparametric models are specified by a parameter vector, $\boldsymbol{\theta}$, and the data, $\mathcal{D} = \{\mathbf{x}_i, y_i\}_{i=1}^{N}$. Equation (0.53) reveals this data dependency explicitly, since the data, $\mathbf{x}_i$, appear on the right hand side of equation (0.53) inside the kernel function, $\kappa$. Because all computations of equation (0.53) are performed on $\mathbf{x}$, the stimulus space, this representation of kernel regression models can be evaluated efficiently. Nowhere is the explicit computation of the potentially very large stimulus feature matrix, $\Phi(\mathbf{x})$, required.

**6 The effects of noise distribution, loss functional, and likelihood**

The third constituent of the MAP framework is the noise distribution. This probability distribution characterizes the model residuals that are not explained by the models within the assumed model class. Gaussian noise is the most commonly assumed noise distribution, and it is the only noise distribution we have considered so far. However, the Gaussian noise assumption is often not valid for neuronal response data. For example, the response distribution is often positively skewed for many continuous response variables, such as local field potential or instantaneous firing rate. The Gaussian noise distribution is symmetric about its mean, so it is not efficient at describing such response asymmetry. Moreover, the response variable may be discrete. For example, the response may be measured as a spike count, or binary variable, indicating whether there is a spike or not. These discrete response variables are poorly modeled by the Gaussian noise distribution.

When a wrong noise distribution is assumed, two scenarios may happen. First, if the assumed noise distribution is too broad, noise variability may be admitted as true system response. Consequently, the model will tend to overfit the noise, and become highly biased by outliers in the data. Second, if the assumed noise distribution is too restrictive, then true response variability may be mistaken as noise. The result is that the model will underfit the data, and it will not be able to explain all the response variability. In practice, the assumed noise distribution is often too broad in certain regions, and too restrictive elsewhere. In general, the estimation bias is often difficult to quantify, because it is difficult to distinguish random noise from the response of a stochastic system.

In this section, we will examine how the assumed noise distribution influences the MAP estimate of the model parameters. We will start our exploration by deriving the MAP objective of simple algorithms under different noise assumptions. For the more complex algorithms, we will present the algorithms first. Then their MAP objective will be derived to reveal the underlying noise distribution. In some cases, the noise distribution may not have a closed form. In other cases, we may only get as far as revealing the likelihood function rather than the actual noise distribution itself.

*6.1 Discrete noise distributions and exponential family noise distributions*

The responses of neuronal systems are mediated by a change of the membrane potentials. Because complex biophysical processes generate these voltage changes, it is often difficult to characterize the noise without a detail biophysical model of these systems. Consequently, standard statistical distributions are often used to approximate the true underlying noise distribution. In statistics, there is a versatile family of parametric distributions that can approximate many univariate distributions arising from physical processes. This family is known as the *exponential family*, and it includes many well known distributions: Gaussian, Poisson, Bernoulli, gamma, binomial, and many others. Because the exponential family includes both continuous and discrete distributions, it is well suited for approximating the noise distribution in neuronal systems. We have already seen many examples of Gaussian noise. In the following sections, we will examine two commonly used distributions of the exponential family for modeling neuronal noise: Poisson (§6.1.1) and Bernoulli (§6.1.2). Then we will discuss the theory for a general exponential family noise distribution in section §6.1.3. Under these noise assumptions, the MAP estimate usually does not have a closed form solution. Therefore, we will only derive the MAP objective function, $g_{MAP}(\theta)$. The MAP estimate of $\theta$ must be computed numerically by minimizing $g_{MAP}(\theta)$ via optimization algorithms.

6.1.1 Poisson noise distribution and count regression

In neuronal SI, the response variable often records the spike counts, $y_i$, elicited by a particular stimulus $\mathbf{x}_i$. Poisson distribution models the variability in spike counts better than the Gaussian distribution. Therefore it is appropriate to assume a Poisson noise distribution for data sets that contains spike count response. The regression algorithms that assume Poisson noise are referred to as *count regression*.

Because the Poisson distribution is $p_{Pois}(y|\mu) \equiv e^{-\mu}\mu^y/y!$, where $\mu$ is the Poisson mean with $\mu_i = f_\theta(x_i)$, the likelihood under the Poisson noise assumption can be written as

$$P(Y|X,\theta) = p(y_i | f_\theta(x_i)) = \prod_{i=1}^{N} \frac{e^{-f_\theta(x_i)}[f_\theta(x_i)]^{y_i}}{y_i!}. \tag{0.54}$$

Taking the negative log of the likelihood (0.54) gives the loss functional under the Poisson noise assumption. So the MAP objective can be written as

$$g_{MAP}(\theta) = -\sum_{i=1}^{N} \log(p(y_i | f_\theta(x_i))) - \log(p(\theta))$$

$$= \sum_{i=1}^{N} \{f_\theta(x_i) - y_i \log[f_\theta(x_i)] + \log(y_i!)\} - \log p(\theta). \tag{0.55}$$

Because $\sum_{i=1}^{N} \log(y_i!)$ is not a function of the parameter vector, $\theta$, it can be treated as a constant that does not affect the minimum of equation (0.55). Therefore the MAP objective (0.55) for count regression can be simplified as

$$g_{MAP}(\theta) = \sum_{i=1}^{N} \{f_\theta(x_i) - y_i \log[f_\theta(x_i)]\} - \log p(\theta). \tag{0.56}$$

Note that equation (0.56) contains a log function, which is undefined for non-positive arguments. To ensure the MAP objective (0.56) is well defined, the argument of the log function must be positive, so $f_\theta(x)$ must be a positive function of the stimulus. Thus, an admissible model class for count regression must produce strictly positive responses. Because a linear model can potentially produce negative response, it is not an admissible model for count regression, but a positively rectified linear model is admissible. After choosing an admissible model class, the remaining quantity that needs to be specified is the prior distributions, $p(\boldsymbol{\theta})$. In theory, any prior presented in section §3 may be used. By choosing a different $p(\boldsymbol{\theta})$, it is possible to derive various forms of regularized count regression.

### 6.1.2 Bernoulli noise distribution and binary regression

When the spike count response is recorded at a temporal resolution that is finer than the refractory period of the neuronal system, the response becomes an *indicator variable*. This indicator variable is a type of binary variable that marks the presence or absence of a spike with 1's and 0's respectively. For such binary response data, the Bernoulli distribution provides a even better description of the response variability than the Poisson distribution. Therefore a Bernoulli noise distribution may be assumed when analyzing data sets that contain binary response variables. The regression algorithms that assumed Bernoulli noise are referred to as *binary regression*.

Because the Bernoulli distribution is $p_{Ber}(y|\mu) \equiv \mu^{y}(1-\mu)^{1-y}$, where $\mu$ is its mean, the likelihood under the Bernoulli noise assumption may be written as

$$\mathcal{P}(\mathbf{Y}|\mathbf{X},\boldsymbol{\theta}) = p(y_i | f_{\boldsymbol{\theta}}(\mathbf{x}_i)) = \prod_{i=1}^{N} f_{\boldsymbol{\theta}}(\mathbf{x}_i)^{y_i} \left[1 - f_{\boldsymbol{\theta}}(\mathbf{x}_i)\right]^{1-y_i}. \qquad (0.57)$$

Taking the negative log of the likelihood (0.57) gives the loss functional under the Bernoulli noise assumption. The MAP objective can then be derived as:

$$g_{MAP}(\boldsymbol{\theta}) = \sum_{i=1}^{N} \left\{ -y_i \log\left[f_{\boldsymbol{\theta}}(\mathbf{x}_i)\right] - (1-y_i)\log\left[1 - f_{\boldsymbol{\theta}}(\mathbf{x}_i)\right] \right\} - \log p(\boldsymbol{\theta})$$

$$= \sum_{i=1}^{N} \left\{ -y_i \log \frac{f_\theta(\mathbf{x}_i)}{1-f_\theta(\mathbf{x}_i)} - \log\left[1-f_\theta(\mathbf{x}_i)\right] \right\} - \log p(\theta). \qquad (0.58)$$

Because the mean, $\mu$, of the Bernoulli distribution represents the probability $p_{Ber}(y=1)$, it can only take values between 0 and 1. Therefore the admissible model classes for binary regression must have range within $[0,+1]$. After choosing an admissible model class, it is still necessary to choose a prior before we can minimize the MAP objective (0.58). Like count regression, any priors presented in section §3 may be used, and different forms of regularized binary regression may be derived by choosing a different prior.

### 6.1.3 Exponential family noise distributions and generalized linear models

As alluded earlier in section §6.1, the exponential family provides a large family of parametric distributions that can model the unknown noise distribution of the neuronal system. We have already seen three examples of noise distribution from the exponential family: Gaussian, Poisson, and Bernoulli. When these three distributions are insufficient for characterizing the noise variability, it is advisable to seek surrogate noise distributions in the exponential family. Therefore it is useful to derive the MAP objective under a general exponential family noise distribution. Because this section is a generalization of the previous two sections, we will often refer to sections §6.1.1 and §6.1.2 for examples.

The distributions in the exponential family have the general form of

$$p_{EF}(y|\mu) \equiv \exp\left[y \cdot a(\mu) + b(\mu) + c(y)\right], \qquad (0.59)$$

where $a(.)$, $b(.)$ and $c(.)$ are known functions (Dobson, 2002). When the noise distribution is assumed to be a member of the exponential family, the likelihood can be written as

$$P(\mathbf{Y}|\mathbf{X},\boldsymbol{\theta}) = p(y_i | f_{\boldsymbol{\theta}}(\mathbf{x}_i)) = \prod_{i=1}^{N} \exp\{y_i a[f_{\boldsymbol{\theta}}(\mathbf{x}_i)] + b[f_{\boldsymbol{\theta}}(\mathbf{x}_i)] + c[y_i]\}. \quad (0.60)$$

The loss functional for a general noise distribution from the exponential family may be obtained by taking the negative log of the likelihood in equation (0.60). Because $c[y_i]$ is independent of $\boldsymbol{\theta}$, it can be ignored, so the MAP objective can be written as

$$g_{MAP}(\boldsymbol{\theta}) = \sum_{i=1}^{N} \{-y_i a[f_{\boldsymbol{\theta}}(\mathbf{x}_i)] - b[f_{\boldsymbol{\theta}}(\mathbf{x}_i)]\} - \log p(\boldsymbol{\theta}). \quad (0.61)$$

Because equation (0.61) is the MAP objective for any noise distribution in the exponential family, the MAP objective under Poisson and Bernoulli noise must be derivable from equation (0.61). Indeed, when Poisson noise is assumed, the count regression MAP objective (0.56) is obtained by setting, $a(\mu) = \log(\mu)$ and $b(\mu) = -\mu$. When Bernoulli noise is assumed, the binary regression MAP objective (0.58) is obtained by setting $a(\mu) = \text{logit}(\mu)$, where $\text{logit}(\mu) \equiv \log[\mu/(1-\mu)]$, and $b(\mu) = \log(1-\mu)$.

Note that the mapping function, $f_{\boldsymbol{\theta}}(\mathbf{x})$, appears in the MAP objective (0.61) as arguments to the functions $a(.)$ and $b(.)$, which may not be defined for all values that $f_{\boldsymbol{\theta}}(\mathbf{x})$ may produce. To ensure the MAP objective is well-defined, the range of $f_{\boldsymbol{\theta}}(\mathbf{x})$ must lie within regions where $a(.)$ and $b(.)$ are defined. This constraint will determine the admissible model classes for regression under exponential family noise. After choosing an admissible model class and a suitable prior, the MAP objective may be optimized to obtain the MAP estimate. However, not all admissible models lead to a convex MAP objective. Therefore equation (0.61) may be plagued with local minima, and its minimization may still be difficult.

To ensure the existence of a unique minimum for the MAP objective (0.61), it is necessary to have a model class that is admissible and makes equation (0.61) convex. Such model class exists and it is referred to as the *generalized linear model* (GLM). The models in GLM assume a mapping function of the form $f_\theta(\mathbf{x}) = a^{-1}(\mathbf{x}^T\boldsymbol{\theta})$, where $a(.)$ is a monotonic nonlinear function known as the *link function*. Because $a(.)$ is also used to define the distributions in the exponential family via equation (0.59), there is one GLM for every noise distribution from the exponential family. For example, the GLM for Poisson noise uses an exponential link function,

$$f_\theta(\mathbf{x}) = \log^{-1}(\mathbf{x}^T\boldsymbol{\theta}) = \exp(\mathbf{x}^T\boldsymbol{\theta}). \tag{0.62}$$

Clearly equation (0.62) is positive. Therefore it is an admissible model for count regression. The GLM in (0.62) is a special kind of count regression known as *Poisson regression*. The GLM for Bernoulli noise uses the logistic link function,

$$f_\theta(\mathbf{x}) = \text{logit}^{-1}(\mathbf{x}^T\boldsymbol{\theta}) = \frac{1}{1+\exp(-\mathbf{x}^T\boldsymbol{\theta})}. \tag{0.63}$$

It can be shown that equation (0.63) is bounded in the range $[0,+1]$, so it defines an admissible model for binary regression. The GLM in (0.63) is a special kind of binary regression known as *logistic regression*. In general, when using GLM, the loss functional in (0.61) is automatically a convex function of the model parameters, $\boldsymbol{\theta}$. Therefore, a unique ML estimate exists, and it can be obtained by *iterative weighted least-squares* (Dobson, 2002).

After choosing an appropriate GLM for the assumed exponential family noise distribution, it is necessary to assume a prior to completely specify the MAP objective (0.61). Because Gaussian priors induce regularizers that are quadratic functions of $\boldsymbol{\theta}$, these regularizers are also convex functions of $\boldsymbol{\theta}$. Since the sum of two convex functions is convex, the MAP objective (0.61) for a GLM under a Gaussian prior will certainly be convex. Thus, a unique MAP estimate is guaranteed, and the MAP estimate may be computed by gradient-based optimization algorithms.

*6.2 Implicitly assumed noise distributions and support vector regression*

Recall that system identification algorithms can be derived from the optimization perspective without making any reference to the underlying noise distributions. The underlying noise distributions for these algorithms are implicitly assumed by using a particular form of loss functional. When the noise mechanism that generated the data is not well characterized, it is computationally advantageous to derive the MAP objective starting from the loss functional. Assuming an incorrect noise distribution not only biases the MAP estimate, it can also make the MAP objective difficult to minimize. Choosing a loss functional that is easy to minimize could at least simplify MAP computation. Rather than starting from the noise distribution, this section will analyze system identification algorithms starting from their loss functional. These algorithms include SVR and other forms of robust regression.

SVR, like kRR (§5.1), is a kernel regression algorithm developed in the machine learning community. However, SVR differs from kRR because it does not assume a Gaussian noise distribution, so it does not use the familiar *square loss* (Figure 4a). Instead, it uses Vapnik's $\varepsilon$-*insensitive loss* [(Figure 4b), (Vapnik, 1995)], defined by

$$|\delta|_\varepsilon \equiv \max\left(0, |\delta| - \varepsilon\right) = \begin{cases} 0 & \text{if } |\delta| \leq \varepsilon \\ |\delta| - \varepsilon & \text{otherwise}. \end{cases} \tag{0.64}$$

Therefore the assumed noise distribution for SVR must be non-Gaussian. It can be shown that the underlying noise distribution for the $\varepsilon$-insensitive loss is an additive mixture of Gaussian noise with random mean and variance (Pontil and Verri, 1998).

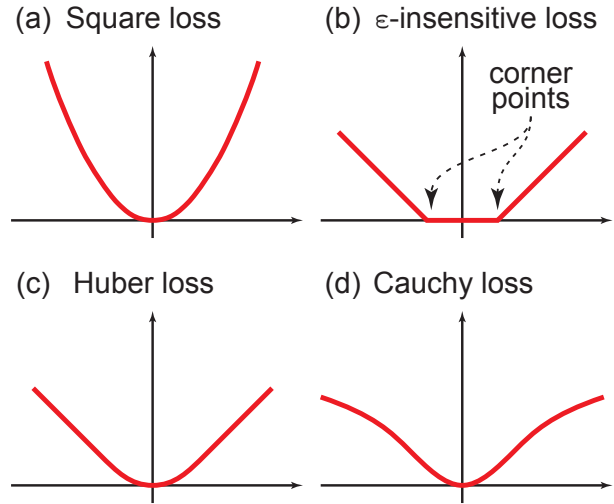

**Figure 4: Loss Functionals.**

Except for the noise distribution, the underlying model class and prior for SVR are the same as kRR. Therefore the MAP objective for SVR can be derived simply by replacing the square loss in equation (0.51) by the $\varepsilon$-insensitive loss,

$$g_{MAP}(\boldsymbol{\beta}) = \sum_{i=1}^{N} \left| y_i - \Phi(\mathbf{x}_i)^T \boldsymbol{\beta} \right|_{\varepsilon} + \lambda \|\boldsymbol{\beta}\|_2^2. \tag{0.65}$$

Now, it is clear that SVR is nothing more than a $\ell_2$-norm regularized linear regression in a feature space where the $\varepsilon$-insensitive loss is used to measure model failure. Because SVR shares many common features with kRR, it is advisable to revisit section §5.1.1 and §5.1.2 for discussions concerning the regularization, representation, and evaluation of SVR.

Note that the defining feature that distinguishes SVR from other neuronal system identification algorithms is the use of the $\varepsilon$-insensitive loss. This loss is a convex function. This makes the SVR objective (0.65) convex. So SVR always has a unique MAP estimate. However, because the $\varepsilon$-insensitive loss has corner points (Figure 4b), the SVR objective is not smooth. Therefore the MAP estimate for SVR cannot be obtained via gradient-based algorithms. In practice, the MAP estimate is obtained by *quadratic programming*, which minimizes the quadratic regularizer,

$$\boldsymbol{\beta}^* = \arg\min_{\beta}\left[\lambda\|\boldsymbol{\beta}\|_2^2 + \sum_{i=1}^{N}\left(\xi_i^+ + \xi_i^-\right)\right],$$

subject to linear constraints imposed by the loss functional:

$$\varepsilon + \xi_i^+ - y_i + \langle\Phi(\mathbf{x}_i),\boldsymbol{\beta}\rangle \geq 0$$
$$\varepsilon + \xi_i^- - \langle\Phi(\mathbf{x}_i),\boldsymbol{\beta}\rangle + y_i \geq 0$$
$$\xi_i^+, \xi_i^- \geq 0.$$

This differs from most neuronal system identification algorithms, which simultaneously minimizes the loss functional and the regularizers.

The corner points in the SVR objective (0.65) introduced by the $\varepsilon$-insensitive loss will also make the MAP estimate of SVR sparse. This phenomenon is similar to the case when the spherical Laplacian prior is assumed (§3.8). The difference here is that the corner points in the SVR objective (0.65) are introduced by the loss functional, whereas the corner points in the lasso objective (0.32) are introduced by the regularizer. However, to an optimization algorithm, there is no distinction between the loss functional and the regularizer. The optimization algorithm only sees the values of the MAP objective. As long as the MAP objective contains corner points, the MAP estimate will be sparse.

6.2.1 Using different robust loss functionals in regression algorithms

The $\varepsilon$-insensitive loss was originally used in SVR as an approximation to *Huber's loss*, which incurs a quadratic penalty for small residuals while penalizing large residuals linearly (Figure 4c). When the underlying noise distribution is not well-characterized, theoretical work in robust statistics suggest using a noise distribution that can account for the presence of outliers (Huber, 1972). In practice, this amounts to assuming a noise distribution that has heavier tails than the Gaussian. Consequently, the induced loss functional is a sub-quadratic *robust loss functional*.

Many robust loss functionals may be used in a regression algorithm. Both the Huber and the linear $\varepsilon$-insensitive loss are convex losses, but there are also many non-convex robust loss functionals. These include the Cauchy loss and the inverted Gaussian loss, which are even more robust and resistant to outliers than the convex losses (Figure 4d). Depending on the optimization algorithm used to minimize the MAP objective and the model class, some loss functionals may be more practical than others. For example, if gradient-based optimization is used, the smooth Huber's loss may be more appropriate than the non-smooth $\varepsilon$-insensitive loss. But when constraint optimization algorithms are used, the piecewise linear $\varepsilon$-insensitive loss offers computational advantages over the smooth Huber's loss. This is because the loss functional can be treated as a set of constraints as in SVR. And having a piecewise linear loss make it very easy to specify these constraints, because all the constraints will be linear. For model classes that are non-convex in the model parameters, such as artificial neural networks, the uniqueness of the extremum cannot be guaranteed regardless of the convexity of the loss functional. In these situations, non-convex losses should be used, if they offer more robustness than the convex losses.

*6.3 Biophysical noise from leaky integrate-and-fire spiking mechanism*

Up until now, the MAP objectives for neuronal system identification algorithms have been derived under the assumption that the data samples are independent. This allows the likelihood to be written as a product of the noise distribution evaluated at the data samples. While this independence assumption is computationally convenient, it is not necessarily biophysically plausible. Realistic neuronal responses are driven by natural stimuli, which are temporally correlated; so the response must have temporal correlation. Because neurons have refractory periods, the response data cannot be completely independent. One simple model that can capture these dependencies in neuronal responses is the leaky integrate-and-fire (LIF) model (Gerstner and Kistler, 2002). This model may be used as an output nonlinearity to model the temporal dependencies of the response in a neuronal system identification algorithm.

A neuronal system identification algorithm that uses the LIF model as output nonlinearity is referred to as a linear-nonlinear-LIF (LN-LIF) model (Pillow and Simoncelli, 2003). The LN-LIF model can be viewed as a *latent variable model* in statistics. In this model, an unobserved latent variable, the voltage, $V$, will describe the temporal dependence between the responses at two different times. And the dynamics of the voltage is governed by the differential equation

$$\tau \frac{dV}{dt} = -(V - V_L) + RI_{in}(t) + \sigma \mathcal{N}_t. \qquad (0.66)$$

Here $\tau$ is a decay time constant, $V_L$ is leak reversal potential, $R$ is a constant representing the membrane resistance, $\mathcal{N}_t$ is a zero-mean unit-variance Gaussian noise process, and $\sigma$ is a scale parameter. The driving force in equation (0.66) is the input current, given by the linear model $I_{in}(t) = \mathbf{x}(t)^T \boldsymbol{\beta}$. Because the voltage in equation (0.66) is never observed, it can be conveniently scaled to the range $[V_L = 0, V_{th} = 1]$, where $V_{th}$ is the spiking threshold (Pillow et al., 2003). In the LIF model, a spiking event will always reset the voltage to the resting state, $V_r$. So whenever a spike occurs, the dynamics of the neuronal system are completely renewed. Therefore the dynamics of the voltage at any time only depends on the voltages up to the time of the previous spike.

In the absence of noise, the evolution of voltage is given by the solution to equation (0.66) with $\sigma = 0$,

$$V(t) = V_r e^{-(t-t_i)/\tau} + R \int_{t_i}^{t} \left[ \mathbf{x}(s)^T \boldsymbol{\beta} \right] e^{-(t-s)/\tau} ds. \qquad (0.67)$$

Because of the renewal property at the time of spiking, the voltage dynamics between each inter-spike interval (ISI) are independent, conditioned upon the observed spikes. It can be shown (Appendix C) that in the presence of Gaussian noise, the likelihood for the LN-LIF model can be written as a product of Gaussian integrals,

$$\mathcal{P}(\mathbf{Y} \mid \mathbf{X}, \boldsymbol{\theta}) = \prod_{j=1}^{m} \int dV_{\mathbf{T}_j} \mathcal{N}_j \left[ V_{\mathbf{T}_j}, \Sigma_{\mathbf{T}_j} \mid \mathbf{x}_{\mathbf{T}_j}, \boldsymbol{\theta} \right]. \qquad (0.68)$$

Here, $\mathbf{T}_j$ denotes all samples (or time bins) that falls within the $j$th ISI, and $\Sigma_{\mathbf{T}_j}$ is the covariance matrix of the samples within the $j$th ISI. Because the likelihood in (0.68) is log-convex, efficient gradient-based algorithms have been developed for optimizing this likelihood (Paninski et al., 2004). The model parameters of the LN-LIF model, $\boldsymbol{\theta} = \{\boldsymbol{\beta}, \tau, V_r, \sigma\}$, can thus be obtained through ML estimation by maximizing equation (0.68). Because the ML estimate of $\boldsymbol{\theta}$ is merely the MAP estimate under the flat prior, the LN-LIF model fits nicely under the MAP framework. To derive a regularized version of the LN-LIF model, it is necessary to assume a prior over the model parameters. As long as the regularizer induced by the assumed prior is also log-convex, a unique MAP estimate for the model parameters is theoretically guaranteed.

### 6.3.1 Modeling the effects of after-currents

In realistic neuronal systems, the sub-threshold dynamics of the voltage is modulated by an important biophysical process, known as the after-current (Stauffer et al., 2007). An after-current, $I_a(t)$, is a current-waveform of fixed size and shape that is generated when an action potential is fired. To include the effect of after-currents, an extension of the LN-LIF model has been developed (Paninski et al., 2004; Pillow et al., 2003). In this model, the after-current, $I_a(t)$, is added to the input current wherever a spiking event occurs. So the total input current is

$$I_{in}(t) = \mathbf{x}(t)^T \boldsymbol{\beta} + \sum_{j=1}^{n \ni t \geq t_n} I_a(t - t_j).$$

Here $t_1, t_2, \cdots, t_n$ are the spike times that precede the present time $t$. Including the after-current will increase the number of model parameters, and this will increase the data requirement for this extended LN-LIF model. The parameter vector of this model is $\boldsymbol{\theta} = \{\boldsymbol{\beta}, I_a(t), \tau, V_r, \sigma\}$, where $I_a(t)$ is treated as a vector whose length equals the length of the after-current waveform. When sufficiently large data sets are available, the linear kernel, $\boldsymbol{\beta}$, of the extended LN-LIF model can be recovered (Paninski et al., 2004; Pillow et al., 2003). When data is limited, however, assuming a suitable prior can mitigate the effect of data limitation. Because the extended LN-LIF model has several different groups of model parameters, different priors may be assumed for the different parameter groups. For example, a smooth prior may be assumed for $I_a(t)$, whereas a spherical Gaussian prior may be assumed for $\boldsymbol{\beta}$.

*6.4 Nonparametric noise and maximally informative dimensions*

In deriving the MAP objective of an system identification algorithm, it is necessary to specify all three MAP constituents. The noise distribution is often specified directly by assuming a parametric form. It is also possible to specify the loss functional, which implicitly defines the assumed noise distribution parametrically. In either case, because the likelihood of the model parameters is a function of the noise distribution evaluated at the data samples, the likelihood is also a parametric distribution. However, because the likelihood is a probability distribution, it can also be approximated nonparametrically. Maximally informative dimensions (MID) is a system identification algorithm that does not assume a parametric form for the likelihood. Therefore the underlying noise distribution must also be nonparametric.

The method of maximally informative dimensions (MID) was developed as an extension of STA (§3.2) to correct for the bias introduced by the non-Gaussian statistics of natural stimuli. It has been shown that this bias is only apparent when nonlinear model classes are used (Sharpee et al., 2004). MID seeks a relevant subspace of the stimulus ensemble that maximizes the mutual information between the subspace projected stimulus and the response. Because mutual information is a nonparametric measure between two stochastic variables, the underlying noise distribution of MID is nonparametric. The effectiveness of MID has been demonstrated in visual system identification (Sharpee et al., 2004, 2006).

The maximally informative dimension, $\boldsymbol{\beta}_1$, is obtained by maximizing the mutual information, $I(\mathbf{x}^T\boldsymbol{\beta}, y)$, or equivalently minimizing $-I(\mathbf{x}^T\boldsymbol{\beta}, y)$. Using the definition of mutual information, we can write the MID objective function as

$$g_{MID}(\boldsymbol{\beta}) = -I(\mathbf{x}^T\boldsymbol{\beta}, y) \equiv -\sum_{\mathbf{x}^T\boldsymbol{\beta}} \sum_y p(\mathbf{x}^T\boldsymbol{\beta}, y) \log \frac{p(\mathbf{x}^T\boldsymbol{\beta}, y)}{p(\mathbf{x}^T\boldsymbol{\beta})p(y)}. \quad (0.69)$$

Equation (0.69) can be simplified by factoring $p(\mathbf{x}^T\boldsymbol{\beta}, y)$ as $p(\mathbf{x}^T\boldsymbol{\beta}|y)p(y)$, so

$$g_{MID}(\boldsymbol{\beta}) = -\sum_y p(y) \left[ \sum_{\mathbf{x}^T\boldsymbol{\beta}} p(\mathbf{x}^T\boldsymbol{\beta}|y) \log \frac{p(\mathbf{x}^T\boldsymbol{\beta}|y)}{p(\mathbf{x}^T\boldsymbol{\beta})} \right]. \quad (0.70)$$

The quantity inside the square bracket of equation (0.70) is the Kullback-Leibler (KL) divergence between $p(\mathbf{x}^T\boldsymbol{\beta}|y)$ and $p(\mathbf{x}^T\boldsymbol{\beta})$. Recall that the KL divergence, $D_{KL}$, between two distributions, $p_1(z)$ and $p_2(z)$, is defined by

$$D_{KL}\left[p_1(z)\|p_2(z)\right] \equiv \sum_z p_1(z) \log \frac{p_1(z)}{p_2(z)}.$$

Therefore equation (0.70) may be written equivalently as

$$g_{MID}(\boldsymbol{\beta}) = -\sum_y p(y) D_{KL}\left[p(\mathbf{x}^T\boldsymbol{\beta}|y)\|p(\mathbf{x}^T\boldsymbol{\beta})\right]. \quad (0.71)$$

It has been argued that the most informative stimulus dimension can be obtained through the following minimization problem,

$$\boldsymbol{\beta}_1 = \arg\min_{\boldsymbol{\beta}} -D_{KL}\left[p(\mathbf{x}^T\boldsymbol{\beta}|y)\|p(\mathbf{x}^T\boldsymbol{\beta})\right]. \quad (0.72)$$

The argument is that the only part of the MID objective, $g_{MID}(\boldsymbol{\beta})$, that depends on $\boldsymbol{\beta}$ is in the KL divergence. The outer summation in equation (0.71) is an expectation over $y$, and should not affect the minimum of $g_{MID}(\boldsymbol{\beta})$. However, because the KL divergence also has dependence on $y$, it can be shown that the most informative dimension obtained via equation (0.72) is biased. To obtain the true maximally informative dimension, it is necessary to optimize equation (0.71).

Although the MID objective (0.69) looks nothing like the MAP objective in equation (0.4), it can be shown (Appendix D) that in the limit of large sample size,

$$g_{MAP}(\boldsymbol{\beta}) = N \cdot g_{MID}(\boldsymbol{\beta}) - \sum_{i=1}^{N} \log p(y_i) - \log p(\boldsymbol{\beta}).  \quad (0.73)$$

The second term in equation (0.73) is independent of $\boldsymbol{\beta}$ and has no effect on the location of the extremum. If the prior is flat (independent of $\boldsymbol{\beta}$), then the extremum of the MAP objective will coincide with the extremum of the MID objective. Thus, in the limit of large sample size, the MID estimate of the model parameter, $\boldsymbol{\beta}_1$, is identical to the MAP estimate, $\boldsymbol{\beta}^*$, under the flat prior.

Recall the crucial difference that distinguishes MID from other neuronal system identification algorithms is the fact that MID does not assume a parametric form for the noise distribution. Therefore, all probability distributions used in the computation of MID are estimated directly from the data. The standard implementation of MID uses the normalized histograms of $p(\mathbf{x}^T\boldsymbol{\beta}|y)$, $p(\mathbf{x}^T\boldsymbol{\beta})$, and $p(y)$ to approximate the corresponding probability distributions. In this implementation, the bin-size (or the number of bins) of the histogram can be viewed as a hyperparameter that must be determined by the user. Therefore bin-size selection can be view as a form of regularization, since it controls the bumpiness of the histograms used to compute the mutual information. Because there is only one hyperparameter, the optimal bin-size can be determined by cross-validation.

# 7 Conclusion

The MAP framework provides a single inferential view for all popular neuronal system identification algorithms. These algorithms include: STA (§3.2), normalized reverse correlation (§3.2), ridge regression (§3.3), linear regression with ARD (§3.4), linear regression under the smooth prior (§3.7), boosting (§3.8), linearized reverse correlation (§4.1), LSRC (§4.1.1), second-order Wiener-Volterra models (§4.2), STC (§4.2.2), kRR (§5.1), SVR (§6.2), the LN-LIF model (§6.3), and MID (§6.4). This framework is very general and appears to be universally encompassing. It includes simple statistical heuristics to algorithms that incorporate very complex and realistic biophysical processes. The hidden assumptions build into each algorithm are made explicit in the form of three MAP constituents: the model class, the noise distribution and the prior. Understanding these constituents can aid the choice of the most appropriate algorithms for a particular data set.

The MAP framework also offers an optimization perspective on neuronal system identification through the MAP objective function, loss functional, and the regularizer. This dual perspective links the theoretical analysis of the neuronal mapping function with the actual implementation and computation of the neuronal system identification algorithm. Understanding the optimization perspective not only shows the strength and limitations of an algorithm, it also reveals whether a particular algorithm is computationally tractable.

Most importantly, the MAP framework can facilitate the development of novel neuronal system identification algorithms. This can be achieved by incorporating realistic assumptions into the priors and the noise distributions. We have already seen some examples of biophysically plausible priors in sections §3.6 to §3.8, and some realistic noise distributions in sections §6.1.1, §6.1.2, and §6.3. Because the three MAP constituents are independent, recombining the existing MAP constituents in new ways can derive novel algorithms for neuronal SI.

# Appendix

**Appendix A: The effective posterior distribution of the model**

In this appendix, we will derive the effective posterior distribution of the model, conditioned on the observed data. The MAP estimate can then be obtained by maximizing this posterior. We can derive the posterior and the effective posterior distribution (0.2) by factoring the joint distribution (0.1) as a product of the conditional, $\mathcal{P}(f|\mathbf{X},\mathbf{Y})$, and the marginal, $\mathcal{P}(\mathbf{X},\mathbf{Y})$. The posterior of the model is then given by

$$\mathcal{P}(f|\mathbf{X},\mathbf{Y}) = \frac{\mathcal{P}(\mathbf{X},\mathbf{Y},f)}{\mathcal{P}(\mathbf{X},\mathbf{Y})}. \qquad (.74)$$

The denominator in equation ( .74) is called the evidence. It is a marginalization of the joint distribution over all possible models by integrating over $f$, and therefore it has no functional dependence on $f$:

$$\mathcal{P}(\mathbf{X},\mathbf{Y}) = \int df\, \mathcal{P}(\mathbf{X},\mathbf{Y},f)$$
$$= \prod_{i=1}^{N} p(\mathbf{x}_i) \int df \prod_{i=1}^{N} p(y_i|\mathbf{x}_i,f) p(f).$$

Here we have factored the finite product into two separate products, and we have taken the stimulus distribution, $p(\mathbf{x}_i)$, out of the integral because it is not a function of the integrated variable, $f$. The posterior can be further simplified by canceling $p(\mathbf{x}_i)$:

$$\mathcal{P}(f|\mathbf{X},\mathbf{Y}) = \frac{\mathcal{P}(\mathbf{X},\mathbf{Y},f)}{\mathcal{P}(\mathbf{X},\mathbf{Y})} = \frac{\prod_{i=1}^{N} p(y_i|\mathbf{x}_i,f)\, \cancel{p(\mathbf{x}_i)}\, p(f)}{\prod_{i=1}^{N} \cancel{p(\mathbf{x}_i)} \int df \prod_{i=1}^{N} p(y_i|\mathbf{x}_i,f) p(f)}$$

$$= \frac{\prod_{i=1}^{N} p(y_i | \mathbf{x}_i, f) p(f)}{\prod_{i=1}^{N} \int df \prod_{i=1}^{N} p(y_i | \mathbf{x}_i, f) p(f)}. \tag{.75}$$

Because the denominator of this posterior does not depend on $f$, it can be treated as a normalization constant. So the denominator in equation (.75) can be ignored without affecting the extremum of the posterior as a function of $f$. Therefore the effective posterior distribution (0.2) that we need to maximize is given by

$$\mathcal{P}^*(f | \mathbf{X}, \mathbf{Y}) \propto \prod_{i=1}^{N} p(y_i | \mathbf{x}_i, f) p(f).$$

The MAP estimate is the $f$ that maximizes this effective posterior, and it corresponds to the most probable $f$ given the observed data.

∎

**Appendix B: Model representation for kernel regression models**

Because the feature space used in kernel regression algorithms have very high (possibly infinite) dimensionality, kernel regression models are not easily represented. Although the MAP estimate for the model parameters can be derived, and it is symbolically given by

$$\boldsymbol{\beta}^* = \left(\boldsymbol{\Phi}_\mathbf{X}^T \boldsymbol{\Phi}_\mathbf{X} + \lambda \mathbf{I}\right)^{-1} \boldsymbol{\Phi}_\mathbf{X}^T \mathbf{Y}, \tag{.76}$$

computing and storing $\boldsymbol{\beta}^*$ may be numerically intractable. Therefore kernel regression models cannot be evaluated by equation (0.48). In this appendix, we will derive an alternative representation for kernel regression models that can be evaluated.

The key point of this derivation is that the sample size, $N$, is usually much smaller than the feature space dimensionality, $d_\mathcal{F}$, since $d_\mathcal{F}$ can be infinite. Without loss of generality, we will restrict our discussion to the case, where $N \ll d_\mathcal{F}$. In this case, although $\mathbf{\Phi}_\mathbf{X}^T \mathbf{\Phi}_\mathbf{X}$ is a $d_\mathcal{F} \times d_\mathcal{F}$ matrix, it will be rank deficient; and it will be at most rank $N$. Using the push-through identity (Appendix B2) the regularized inverse of the feature autocovariance in equation ( .76) can be written as

$$\left(\mathbf{\Phi}_\mathbf{X}^T \mathbf{\Phi}_\mathbf{X} + \lambda \mathbf{I}\right)^{-1} \mathbf{\Phi}_\mathbf{X}^T = \mathbf{\Phi}_\mathbf{X}^T \left(\mathbf{\Phi}_\mathbf{X} \mathbf{\Phi}_\mathbf{X}^T + \lambda \mathbf{I}\right)^{-1}. \qquad (.77)$$

Substituting equation ( .77) into equation ( .76), the MAP estimate for $\boldsymbol{\beta}$ can be re-expressed as

$$\boldsymbol{\beta}^* = \mathbf{\Phi}_\mathbf{X}^T \left(\mathbf{\Phi}_\mathbf{X} \mathbf{\Phi}_\mathbf{X}^T + \lambda \mathbf{I}\right)^{-1} \mathbf{Y}. \qquad (.78)$$

It might seem that we have not done much, but the matrix $\mathbf{\Phi}_\mathbf{X} \mathbf{\Phi}_\mathbf{X}^T$ is now $N \times N$ rather than $d_\mathcal{F} \times d_\mathcal{F}$. For $N \ll d_\mathcal{F}$, inverting $\left(\mathbf{\Phi}_\mathbf{X} \mathbf{\Phi}_\mathbf{X}^T + \lambda \mathbf{I}\right)^{-1}$ will be much more tractable than inverting $\left(\mathbf{\Phi}_\mathbf{X}^T \mathbf{\Phi}_\mathbf{X} + \lambda \mathbf{I}\right)^{-1}$. However, being able to compute $\left(\mathbf{\Phi}_\mathbf{X} \mathbf{\Phi}_\mathbf{X}^T + \lambda \mathbf{I}\right)^{-1}$ does not solve all the problems, because it is still infeasible to compute the $d_\mathcal{F} \times N$ matrix, $\mathbf{\Phi}_\mathbf{X}^T$.

In order to represent a kernel regression model and evaluate it, we must make use of the kernel trick (Aizerman et al., 1964). This trick converts the computation of an inner product in a feature space into an evaluation of a kernel function in the stimulus space. The kernel function is a symmetric, and positive semi-definite function defined by

$$\kappa\left(\mathbf{x}_i, \mathbf{x}_j\right) \equiv \left\langle \mathbf{\Phi}\left(\mathbf{x}_i\right), \mathbf{\Phi}\left(\mathbf{x}_j\right) \right\rangle. \qquad (.79)$$

Here $\mathbf{x}_i$ and $\mathbf{x}_j$ are arbitrary stimuli from stimulus space that are not necessarily part of our data. The kernel function ( .79) is called the *evaluation functional* of the inner product in the feature space. Because equation ( .79) establishes the equality between the inner product in the feature space and a function evaluation in the stimulus space, we can simply evaluate the function, $\kappa$, to obtain the value of the inner product. This circumvents the explicit transformation of the stimulus into the feature space, which is often impractical if not infeasible. The kernel function is also well defined regardless of the dimensionality of the feature space.

Using the kernel function, it can be shown that $\Phi_X \Phi_X^T$ is a Gram matrix, or equivalently a matrix of kernel evaluations between data samples. Hence we can define the $N \times N$ kernel matrix, $\mathbf{K} \equiv \Phi_X \Phi_X^T$, by $\mathbf{K}_{ij} \equiv \kappa(\mathbf{x}_i, \mathbf{x}_j)$. Now, the only incomputable quantity of equation ( .78) is $\Phi_X^T$.

We will define the computable parts of equation ( .78) as

$$\boldsymbol{\alpha} \equiv (\mathbf{K} + \lambda \mathbf{I})^{-1} \mathbf{Y},$$

where $\boldsymbol{\alpha}$ is a $N \times 1$ vector.

Now, although the MAP estimate is simply $\boldsymbol{\beta}^* = \Phi_X^T \boldsymbol{\alpha}$, it is still inaccessible, because we cannot explicitly compute $\Phi_X^T$. However, all kernel regression model are written as an inner product between $\Phi(\mathbf{x})$ and $\boldsymbol{\beta}^*$. Therefore kernel regression models can also be represented by

$$f_{\boldsymbol{\theta}, \mathcal{D}}(\mathbf{x}) = \langle \Phi(\mathbf{x}), \boldsymbol{\beta}^* \rangle = \langle \Phi(\mathbf{x}), \Phi_X^T \boldsymbol{\alpha} \rangle = \sum_{i=1}^{N} \kappa(\mathbf{x}, \mathbf{x}_i) \alpha_i.$$

Under this representation, kernel regression models are feasibly evaluated regardless of the dimensionality of the feature space.

∎

**Appendix B2: The push-through identity**

The push-through identity states that for any $n \times m$ matrix $\mathbf{A}$ and $m \times n$ matrix $\mathbf{B}$,

$$\mathbf{A}(\lambda \mathbf{I}_m + \mathbf{B}\mathbf{A})^{-1} = (\lambda \mathbf{I}_n + \mathbf{A}\mathbf{B})^{-1} \mathbf{A}.$$

This identity can be proved easily:

$$\mathbf{A}(\lambda\mathbf{I}_m + \mathbf{BA}) = (\lambda\mathbf{I}_n + \mathbf{AB})\mathbf{A}$$

$$\mathbf{A}(\lambda\mathbf{I}_m + \mathbf{BA})(\lambda\mathbf{I}_m + \mathbf{BA})^{-1} = (\lambda\mathbf{I}_n + \mathbf{AB})\mathbf{A}(\lambda\mathbf{I}_m + \mathbf{BA})^{-1}$$

$$\mathbf{A} = (\lambda\mathbf{I}_n + \mathbf{AB})\mathbf{A}(\lambda\mathbf{I}_m + \mathbf{BA})^{-1}$$

$$(\lambda\mathbf{I}_n + \mathbf{AB})^{-1}\mathbf{A} = (\lambda\mathbf{I}_n + \mathbf{AB})^{-1}(\lambda\mathbf{I}_n + \mathbf{AB})\mathbf{A}(\lambda\mathbf{I}_m + \mathbf{BA})^{-1}$$

$$(\lambda\mathbf{I}_n + \mathbf{AB})^{-1}\mathbf{A} = \mathbf{A}(\lambda\mathbf{I}_m + \mathbf{BA})^{-1}$$

∎

**Appendix C: Deriving the likelihood for the LN-LIF model**

To obtain the MAP estimate of the LN-LIF model, it is necessary to have the likelihood and the prior over the parameters of this model. Because the prior may be chosen arbitrarily, we will only derive the likelihood in this appendix. To derive the likelihood for a latent variable model, such as the LN-LIF model, we must have the joint distribution for the data, $\mathbf{X}$ and $\mathbf{Y}$, the model parameters, $\boldsymbol{\theta}$, and the latent variables, $\mathbf{V}$, where $\mathbf{V} = [V_0, V_1, \cdots V_N]^T$. This joint distribution can be factor as

$$\mathcal{P}(\mathbf{X}, \mathbf{Y}, \mathbf{V}, \boldsymbol{\theta}) = \prod_{i=1}^{N} p(y_i | V_i) p(V_i | V_{i-1}, \mathbf{x}_i, \boldsymbol{\theta}) p(\mathbf{x}_i) p(\boldsymbol{\theta}). \quad (.80)$$

Note that the latent voltages, $V_i$, are dependent from one sample to the next, so the response, $y_i$, will also inherit this temporal dependence.

To simplify the joint distribution (.80), we must specify all the probability distributions in the right hand side of equation (.80) that depends on the latent voltage, $V_i$. $p(y_i | V_i)$ can be specified by the assumptions of the LIF model. When the voltage, $V_i$, reaches the spiking threshold, $V_{th} = 1$, a spike is fired, $y_i = 1$, but when the $V_i < V_{th}$, there is no spike, $y_i = 0$.

Because these events occur with complete certainty, they are characterized by delta functions, which have zero uncertainties:

$$p(y_i | V_i) = \begin{cases} \delta(y_i) & V_i < V_{th} \\ \delta(y_i - 1) & V_i \geq V_{th}. \end{cases} \quad (.81)$$

To define $p(V_i | V_{i-1}, \mathbf{x}_i, \boldsymbol{\theta})$ in equation (.80), we make use of an important fact from stochastic differential equations: linear dynamics preserves Gaussianity. Applying this fact to equation (0.66), which governs the dynamics of $V$ within each ISI (when $V_i < V_{th}$), we may conclude that the distribution over $V_i$ must be Gaussian for $V_i < V_{th}$. This is because equation (0.66) is linear in $V$ and has Gaussian noise ($\sigma > 0$). When $V_i$ reaches $V_{th} = 1$, the voltage is reset to the resting state, $V_r$. Because this event always occurs with complete certainty, it is characterized by a delta functions. Therefore the probability of $V_i$ can be written as

$$p(V_i | V_{i-1}, \mathbf{x}_i, \boldsymbol{\theta}) = \begin{cases} \mathcal{N}(V_i, \sigma_i^2) & V_i < V_{th} \\ \delta(V_i - V_r) & V_i \geq V_{th}. \end{cases} \quad (.82)$$

Here $\sigma_i^2$ is the variance of the Gaussian noise in the voltage data at time step $i$, and it is related but generally different from $\sigma$, which is the noise variance of the stochastic process that is in the derivative of $V$.

To obtain the posterior distribution $\mathcal{P}(\boldsymbol{\theta} | \mathbf{X}, \mathbf{Y})$ using techniques in appendix A, we must first marginalize the joint distribution over the unobserved latent variables, $V_i$.

$$\mathcal{P}(\mathbf{X}, \mathbf{Y}, \boldsymbol{\theta}) = \int d\mathbf{V} \mathcal{P}(\mathbf{X}, \mathbf{Y}, \mathbf{V}, \boldsymbol{\theta})$$
$$= \int d\mathbf{V} \prod_{i=1}^{N} p(y_i | V_i) p(V_i | V_{i-1}, \mathbf{x}_i, \boldsymbol{\theta}) p(\mathbf{x}_i) p(\boldsymbol{\theta})$$

Here $d\mathbf{V} = dV_0 dV_1 \cdots dV_N$. Because $V_i$ is conditionally dependent on $V_{i-1}$, the integral over the voltages are linked from one term to another. However, we can split the product in the integral into two parts (one that depends on $V_i$ and one that is independent of $V_i$), and take the product that does not depend on $V_i$ outside the integral over the voltages.

$$P(\mathbf{X},\mathbf{Y},\boldsymbol{\theta}) = \int d\mathbf{V} P(\mathbf{X},\mathbf{Y},\mathbf{V},\boldsymbol{\theta})$$

$$= \int d\mathbf{V} \prod_{i=1}^{N}\left[p(y_i|V_i)p(V_i|V_{i-1},\mathbf{x}_i,\boldsymbol{\theta})\right]\prod_{i=1}^{N}\left[p(\mathbf{x}_i)p(\boldsymbol{\theta})\right]$$

$$= \prod_{i=1}^{N}\left[p(\mathbf{x}_i)p(\boldsymbol{\theta})\right]\int d\mathbf{V}\prod_{i=1}^{N} p(y_i|V_i)p(V_i|V_{i-1},\mathbf{x}_i,\boldsymbol{\theta})$$

Using equation ( .75), one can derive an expression for the posterior over $\boldsymbol{\theta}$:

$$P(\boldsymbol{\theta}|\mathbf{X},\mathbf{Y}) = \frac{P(\mathbf{X},\mathbf{Y},\boldsymbol{\theta})}{P(\mathbf{X},\mathbf{Y})} = \frac{P(\mathbf{X},\mathbf{Y},\boldsymbol{\theta})}{\int d\boldsymbol{\theta} P(\mathbf{X},\mathbf{Y},\boldsymbol{\theta})} = \frac{\int d\mathbf{V} P(\mathbf{X},\mathbf{Y},\mathbf{V},\boldsymbol{\theta})}{\int d\boldsymbol{\theta}\int d\mathbf{V} P(\mathbf{X},\mathbf{Y},\mathbf{V},\boldsymbol{\theta})}$$

$$= \frac{\prod_{i=1}^{N} p(\mathbf{x}_i)p(\boldsymbol{\theta})\int d\mathbf{V}\prod_{i=1}^{N} p(y_i|V_i)p(V_i|V_{i-1},\mathbf{x}_i,\boldsymbol{\theta})}{\int d\boldsymbol{\theta}\prod_{i=1}^{N} p(\mathbf{x}_i)p(\boldsymbol{\theta})\int d\mathbf{V}\prod_{i=1}^{N} p(y_i|V_i)p(V_i|V_{i-1},\mathbf{x}_i,\boldsymbol{\theta})}$$

$$= \frac{\prod_{i=1}^{N} \cancel{p(\mathbf{x}_i)}p(\boldsymbol{\theta})\int d\mathbf{V}\prod_{i=1}^{N} p(y_i|V_i)p(V_i|V_{i-1},\mathbf{x}_i,\boldsymbol{\theta})}{\prod_{i=1}^{N} \cancel{p(\mathbf{x}_i)}\int d\boldsymbol{\theta} p(\boldsymbol{\theta})\int d\mathbf{V}\prod_{i=1}^{N} p(y_i|V_i)p(V_i|V_{i-1},\mathbf{x}_i,\boldsymbol{\theta})}$$

The stimulus dependence term, $\prod_{i=1}^{N} p(\mathbf{x}_i)$, will cancel out as demonstrated in Appendix A. Since the denominator is an integral over $\boldsymbol{\theta}$, it will be independent of $\boldsymbol{\theta}$. Consequently, the denominator will not affect the extremum of the posterior $\mathcal{P}(\boldsymbol{\theta}|\mathbf{X},\mathbf{Y})$ and can be treated as a constant. Hence we can write

$$\mathcal{P}(\boldsymbol{\theta}|\mathbf{X},\mathbf{Y}) \propto p(\boldsymbol{\theta}) \underbrace{\int d\mathbf{V} \prod_{i=1}^{N} p(y_i|V_i) p(V_i|V_{i-1},\mathbf{x}_i,\boldsymbol{\theta})}_{\text{Likelihood}}. \qquad (.83)$$

Clearly $p(\boldsymbol{\theta})$ is the prior, so the remaining multivariate integral must be the likelihood, $\mathcal{L}$. To carry out the multivariate integration in equation (.83), we first integrate over the voltages at the times when a spike is fired, where $p(V_i|V_{i-1},\mathbf{x}_i,\boldsymbol{\theta}) = \delta(V_i - V_r)$. This delta-function will break the chain of conditional dependence, and decouple the integral into segments that correspond to the ISIs,

$$\mathcal{L} = \prod_{j=1}^{m} \left\{ \int \prod_{i=t_{j-1}+1}^{t_j - 1} dV_i \delta(y_i) p(V_i|V_{i-1},\mathbf{x}_i,\boldsymbol{\theta}) \right\}.$$

Here $t_1, t_2, \cdots, t_m$ are the spiking times for the $m$ spikes, and $V_{i-1} = V_r$ whenever $i-1$ equals a spiking time, $t_j$. The delta-function $\delta(y_i)$ came from substituting equation (.81) into equation (.83), since it is certain that there is no spike ($y_i = 0$) within an ISI when $V_i < V_{th}$. However, this delta-function is independent of $\boldsymbol{\theta}$, so it will cancel with the normalization constant of the posterior as $p(\mathbf{x}_i)$ did (Appendix A). From equation (.82), we know that each $p(V_i|V_{i-1},\mathbf{x}_i,\boldsymbol{\theta})$ within an ISI is normally distributed, hence the integrand for each ISI is a product of Gaussian distributions. Since the product of Gaussians is a MVG, the likelihood is a product of MVG integrals

$$\mathcal{L} = \prod_{j=1}^{m} \left\{ \int \prod_{i=t_{j-1}+1}^{t_j-1} dV_i \mathcal{N}\left[V_i, \sigma_i^2(V_{i-1}, \mathbf{x}_i, \boldsymbol{\theta})\right] \right\}$$

$$= \prod_{j=1}^{m} \int dV_{\mathbf{T}_j} \mathcal{N}_j \left[V_{\mathbf{T}_j}, \boldsymbol{\Sigma}_{\mathbf{T}_j} \mid \mathbf{x}_{\mathbf{T}_j}, \boldsymbol{\theta}\right].$$

Here the subscript $\mathbf{T}_j$ represents all samples (or time bins) within the $j$th ISI. And $\mathcal{N}_j$ is MVG, whose covariance matrix, $\boldsymbol{\Sigma}_{\mathbf{T}_j}$, is the covariance matrix of the voltages in $\mathbf{T}_j$.

∎

## Appendix D: The MID objective and the MAP objective

To understand how the method of MID fits into the MAP framework, we need to establish the relationship between the MID objective and the MAP objective. We will show that the MID objective (0.69) is asymptotically equivalent to the MAP objective under a flat prior. We will start by rewrite the MAP objective function (0.4) without changing its minimum. That is, we add and subtract the term $\sum_{i=1}^{N} \log p(y_i)$ and combine one of the added terms with the likelihood,

$$g_{MAP}(\boldsymbol{\beta}) = -\sum_{i=1}^{N} \log \frac{p(y_i \mid \mathbf{x}_i^T \boldsymbol{\beta})}{p(y_i)} - \sum_{i=1}^{N} \log p(y_i) - \log p(\boldsymbol{\beta}). \qquad (.84)$$

Although each $\mathbf{x}_i$ is distinct, $\mathbf{x}^T \boldsymbol{\beta}$ is a projection of high-dimensional stimulus down into a one-dimensional space. Therefore many different $\mathbf{x}_i$ may have similar values of projection. Moreover, MID bins the value of $\mathbf{x}^T \boldsymbol{\beta}$ to create the normalized histogram that approximates $p(\mathbf{x}^T\boldsymbol{\beta})$. Hence equation (.84), can be re-expressed as a double sum over the possible values of $y$ and $\mathbf{x}^T\boldsymbol{\beta}$,

$$g_{MAP}(\boldsymbol{\beta}) = -\sum_{\mathbf{x}^T\boldsymbol{\beta}}\sum_{y} M_{\mathbf{x}^T\boldsymbol{\beta},y} \log\frac{p(y|\mathbf{x}^T\boldsymbol{\beta})}{p(y)} - \sum_{i=1}^{N}\log p(y_i) - \log p(\boldsymbol{\beta}).$$

Here $M_{\mathbf{x}^T\boldsymbol{\beta},y}$ is the counts for observing a particular value of $y$ and $\mathbf{x}^T\boldsymbol{\beta}$. By the law of large number, the frequency count $M_{\mathbf{x}^T\boldsymbol{\beta},y}$ will approximate the joint distribution $p(\mathbf{x}^T\boldsymbol{\beta},y)$ as $N \to \infty$. Therefore the MAP objective can be re-expressed as

$$g_{MAP}(\boldsymbol{\beta}) = -\sum_{\mathbf{x}^T\boldsymbol{\beta}}\sum_{y} Np(\mathbf{x}^T\boldsymbol{\beta},y) \log\frac{p(y|\mathbf{x}^T\boldsymbol{\beta})}{p(y)} - \sum_{i=1}^{N}\log p(y_i) - \log p(\boldsymbol{\beta}). \quad (.85)$$

The first term of equation (.85) can be modified without changing the extremum of the MAP objective by multiplying the numerator and the denominator inside the log by $p(\mathbf{x}^T\boldsymbol{\beta})$. This will convert the conditional distribution, $p(y|\mathbf{x}^T\boldsymbol{\beta})$, inside the log into the joint distribution, $p(y,\mathbf{x}^T\boldsymbol{\beta})$, so the MAP objective can be written as

$$g_{MAP}(\boldsymbol{\beta}) = -N\sum_{\mathbf{x}^T\boldsymbol{\beta}}\sum_{y} p(y,\mathbf{x}^T\boldsymbol{\beta}) \log\frac{p(y,\mathbf{x}^T\boldsymbol{\beta})}{p(y)p(\mathbf{x}^T\boldsymbol{\beta})} - \sum_{i=1}^{N}\log p(y_i) - \log p(\boldsymbol{\beta}). \quad (.86)$$

The first term of equation (.86) is clearly proportional to the mutual information, $I(\mathbf{x}^T\boldsymbol{\beta},y)$, as defined by equation (0.69). Thus in the limit as $N \to \infty$,

$$g_{MAP}(\boldsymbol{\beta}) = N \cdot g_{MID}(\boldsymbol{\beta}) - \sum_{i=1}^{N}\log p(y_i) - \log p(\boldsymbol{\beta}),$$

as we have claimed in equation (0.73).

Sahani, M., and Linden, J.F. (2003). Evidence Optimization Techniques for Estimating Stimulus-Response Functions. In Advances in Neural Information Processing Systems, S. Becker, S. Thrun, and K. Obermayer, eds. (Cambridge, Mass. London: MIT), pp. 301–308.

Sarle, W.S. (1995). Stopped Training and Other Remedies for Overfitting. In Proceedings of the 27th Symposium on the Interface of Computing Science and Statistics, pp. 352–360.

Schölkopf, B., Smola, A., Smola, E., and Müller, K.-R. (1998). Nonlinear Component Analysis as a Kernel Eigenvalue Problem. Neural Comput. *10*, 1299–1319.

Schwartz, O., Chichilnisky, E.J., and Simoncelli, E.P. (2002). Characterizing neural gain control using spike-triggered covariance. In Advances in Neural Information Processing Systems, T.G. Dietterich, S. Becker, and Z. Ghahramani, eds. (Cambridge, Mass. USA: MIT Press), pp. 269–276.

Sen, K., Theunissen, F.E., and Doupe, A.J. (2001). Feature analysis of natural sounds in the songbird auditory forebrain. J. Neurophysiol. *86*, 1445–1458.

Sharpee, T., Rust, N.C., and Bialek, W. (2004). Analyzing neural responses to natural signals: Maximally informative dimensions. Neural Comput. *16*, 223–250.

Sharpee, T.O., Sugihara, H., Kurgansky, A.V., Rebrik, S.P., Stryker, M.P., and Miller, K.D. (2006). Adaptive filtering enhances information transmission in visual cortex. Nature *439*, 936–942.

Simoncelli, E.P., and Heeger, D.J. (1998). A model of neuronal responses in visual area MT. Vision Res. *38*, 743–761.

Simoncelli, E.P., and Olshausen, B.A. (2001). Natural image statistics and neural representation. Annu. Rev. Neurosci. *24*, 1193–1216.

Smyth, D., Willmore, B., Baker, G.E., Thompson, I.D., and Tolhurst, D.J. (2003). The receptive-field organization of simple cells in primary visual cortex of ferrets under natural scene stimulation. J. Neurosci. *23*, 4746–4759.

Snoek, J., Larochelle, H., and Adams, R.P. (2012). Practical Bayesian Optimization of Machine Learning Algorithms. In Advances in Neural Information Processing Systems 25,.

Stanley, G.B. (2002). Adaptive spatiotemporal receptive field estimation in the visual pathway. Neural Comput. *14*, 2925–2946.

Stauffer, E.K., McDonagh, J.C., Hornby, T.G., Reinking, R.M., and Stuart, D.G. (2007). Historical reflections on the afterhyperpolarization--firing rate relation of vertebrate spinal neurons. J. Comp. Physiol. A Neuroethol. Sens. Neural. Behav. Physiol. *193*, 145–158.

Stone, M. (1974). Cross-validatory choice and assessment of statistical predictions. J. R. Stat. Soc. Ser. B Methodol. *36*, 111–147.

95

# Figure Captions:

**Figure 1: Maximum a posteriori (MAP) Inference.** The task of an inference algorithm is to find a function, $f$, in a function space, such that $f$ fits the data, $\{\mathbf{X},\mathbf{Y}\}$, well. Typical function spaces (green ellipse) are too large for this purpose, and they contain many functions that can fit the data perfectly. MAP inference is a way to restrict the size of the inferential space through assumptions on the three MAP constituents, so that a unique function, $f^*$, may be obtained. The three MAP constituents are: (1) the model class, (2) the noise distribution, and (3) the prior. The model class, $\mathcal{M}$, defines a parameterized function class (blue ellipse) over which inference is performed. This strongly restricts the inferential space, because $\mathcal{M}$ contains only functions, $f_{\boldsymbol{\theta}}$, parameterized by the parameter vector, $\boldsymbol{\theta}$. The model class also establishes a one-to-one correspondence between $\boldsymbol{\theta}$ and $f_{\boldsymbol{\theta}}$, so that inference may be performed over the parameter space, $\Theta$, (a vector space) rather than $\mathcal{M}$ (a function space). The noise distribution together with the data defines a probability distribution, the likelihood, over $\mathcal{M}$. The mode of the likelihood is a function, $f_{ML}$, that fits the data best under the assumed noise distribution. The prior defines a subjective distribution over $\mathcal{M}$ (or equivalently over $\Theta$), that specify the plausibility of the functions (or parameters values) in $\mathcal{M}$ (or in $\Theta$). Multiplying the prior by the likelihood gives the posterior distribution. The mode of the posterior distribution is the MAP estimate, $f_{MAP}^*$, that not only fits the data well, but is also plausible under the assumed prior.

**Figure 2: Gaussian priors.** The regions of high prior probability defined by different Gaussian priors are multidimensional ellipsoids (or their degenerate forms) in the parameter space. (a) The spherical Guassian prior is a special case where the parameters are assumed independent with the same prior variance. Therefore, the ellipsoid degenerates to a sphere, because the principal axes all have the same length, and they are aligned to the axes of the parameter space. (b) The independence prior relaxes the constraint that requires all prior variances to be equal, so the principal axes do not have the same length. But the parameters are still independent, so the principal axes are still axes-aligned. (c) The stimulus covariance prior allows correlations in the parameters so the principal axes of the ellipsoid are not axes-aligned. Instead their direction and length are determined by the eigenvectors and eigenvalues of the stimulus autocovariance matrix. (d) The covariance subspace prior is a degenerate case of the stimulus covariance prior, so the directions of the principal axes are again defined by the eigenvectors of the stimulus autocovariance matrix. However, the directions with small prior variances (with eigenvalues $< \varepsilon$) are collapsed to zero, so the ellipsoid lies on a lower dimensional hyperplane. (e) The stimulus subspace prior is another degenerate case of the stimulus covariance prior, so it lies in the same subspace as the covariance subspace prior. However, the principal axes within the subspace have infinite length, so the lower dimensional ellipsoid degenerate into a lower dimensional hyperplane.

**Figure 3: Corner points.** Corner points are locations where the function's derivative does not exist. (a) The regularizer of a 1-dimensional Laplace prior has a corner point at $\beta = 0$. (b) The regularizer of 2-dimensional Laplace prior has corner points all along the axes of the parameter

space. For higher-dimensional Laplace priors, the corner points are on the hyperplanes defined by $\beta_i = 0$.

**Figure 4: Loss Functionals.** Different loss functionals can be used in the MAP objective without considering their underlying noise distribution. If a loss functional is convex, it is guaranteed to have a unique extremum. And if the loss is smooth, then gradient-based optimization algorithm may be used to find the extremum rapidly. (a) The square loss is used in many regression algorithms because it is convex and smooth. However, this loss is sensitive to outliers, because it penalizes large residuals quadratically. (b) The ε-insensitive loss is used in SVR. It is convex, but it is not smooth, due to the two corner points. This loss is more robust to outliers because it penalizes large residuals linearly. (c) The Huber loss is often used in robust regression algorithms because it also penalizes large residuals linearly. Moreover, the Huber loss is convex and smooth. (d) The Cauchy loss is even more robust to outliers, because it penalizes large residuals sub-linearly. Although it is smooth, it is not convex.